\documentclass[reprint, amsmath,amssymb, prl, ]{revtex4-2}
\pdfoutput=1

\usepackage{graphicx}% Include figure files
\usepackage{dcolumn}% Align table columns on decimal point
\usepackage{hyperref} % add hypertext capabilities
\usepackage{mathtools}
\usepackage{siunitx}

\usepackage{graphicx}
\usepackage{float}

\usepackage{color}
\usepackage[dvipsnames]{xcolor}

\begin{document}

\title{The nonlocal dielectric response of water in nanoconfinement}

\author{G. Monet} 
\affiliation{Sorbonne Universit{\'e}, CNRS, Laboratoire de Physique Th{\'e}orique de la Mati{\`e}re Condens{\'e}e (LPTMC, UMR 7600), F-75005 Paris, France}
\author{F. Bresme}
\affiliation{Department of Chemistry, Molecular Sciences Research Hub, Imperial College London,  W12 0BZ 2AZ London, United Kingdom}
\author{A. Kornyshev}
\affiliation{Department of Chemistry, Molecular Sciences Research Hub, Imperial College London,  W12 0BZ 2AZ London, United Kingdom}
	  \author{H. Berthoumieux} 
 \affiliation{Sorbonne Universit{\'e}, CNRS, Laboratoire de Physique Th{\'e}orique de la Mati{\`e}re Condens{\'e}e (LPTMC, UMR 7600), F-75005 Paris, France}

\begin{abstract}
Recent experiments reporting a very low dielectric permittivity for nanoconfined water have renewed the interest to the structure and dielectric properties of water in narrow gaps. Here, we describe such systems with a minimal Landau-Ginzburg field-theory composed of a nonlocal bulk-determined term and a local water-surface interaction term. We show how the interplay between the boundary conditions and intrinsic bulk correlations encodes dielectric properties of confined water. Our theoretical  analysis is supported by molecular dynamics simulations and comparison with the experimental data.
% 595/600
\end{abstract}

\maketitle

% ---------------------------------------------------------------------------------------------- %
{\it Introduction - }
Interest in the dielectric properties of confined water has been boosted by the remarked measurement of the dielectric permittivity of nanometric water layer confined between hydrophobic surfaces \cite{fumagalli2018}. Fumagali et al. reported an anomalously low dielectric constant in the direction perpendicular to the surface. \cite{kalinin2018}
Water permittivity in the vicinity of a surface is inhomogeneous\cite{bonthuis2012,zhang2018} leading to a significant increase of the electrostatic interactions, as postulated in the 1950's by Schellman,\cite{schellman1953} and observed experimentally and in simulations \cite{ballenegger2005,chen2015, sato2018}.
The stability of emulsions and colloidal solutions \cite{bergeron1999,levinger2002}, ion transport and reactivity in  channels of proteins,\cite{gouaux2005}, in subsystems of geological interest \cite{fenter2013} or in nanotechnologic devices \cite{siria2017} are strongly influenced by electrostatic properties of confined water. 
However, a fundamental analytic theory connecting the dielectric response to the properties of the confining surfaces, namely chemical composition, degree and geometry of confinement, is still outstanding.\cite{munozsantiburcio2017} 
At the molecular scale, the relative dielectric permittivity tensor $\epsilon_{\alpha,\beta}(\vec{r}-\vec{r'})$ of bulk water is non local.\cite{kornyshev1986, bopp1998, bopp1996}
The structuration in the fluid at an interface induced by this nonlocality has been widely studied at the atomic scale using molecular dynamics (MD) simulations \cite{vorotyntsev1979,kornyshev1981,kornyshev2007,kornyshev1988,schaaf2016}.
At a coarse-grained scale, continuum nonlocal electrostatics provide a useful framework to quantify the dielectric properties of confined correlated fluids. \cite{kornyshev1981} 
This can be based on phenomenological energy functionals that are written in terms of the polarization field ${\vec m}$. They are the sum of the electrostatic energy depending on the displacement field ${\vec D_0}$ and of a correlation term \cite{hildebrandt2004, maggs2006,berthoumieux2015,vatin2020}. 
It reads
\begin{multline}
	\label{elnonloc}
		\mathcal{U}_{\mathrm{bulk}}[\vec{m},\vec{D}_0] = 
		\frac{1}{2 \epsilon_0} \int{d \vec{r} \left( \vec{D}_0 - \vec{m}(\vec{r}) \right)^2} \\
		+\frac{1}{2 \epsilon_0}\int{d \vec{r} d \vec{r'} m^\alpha(\vec{r}) \mathcal{K}_{\alpha,\beta}(\vec{r},\vec{r'}) m^\beta(\vec{r'})},
\end{multline} 
% Count : +32
where $\epsilon_0$ is the vacuum dielectric permittivity. \par 

We specify the kernel $ \mathcal{K}_{\alpha,\beta}(\vec{r},\vec{r'})$ to mimic the simulated nonlocal dielectric properties of bulk water.  We further introduce a phenomenological interaction energy between the surface and the fluid as a sum of harmonic potentials.
We show that this framework reproduces both MD simulations for two hydrophobic surfaces, graphene and hexagonal boron nitride (hBN), and an experimental data.\cite{fumagalli2018} 
In addition, it formalizes the effect of the confining material on the dielectric properties of 'interfacial water'. \par

\begin{figure}
	\includegraphics{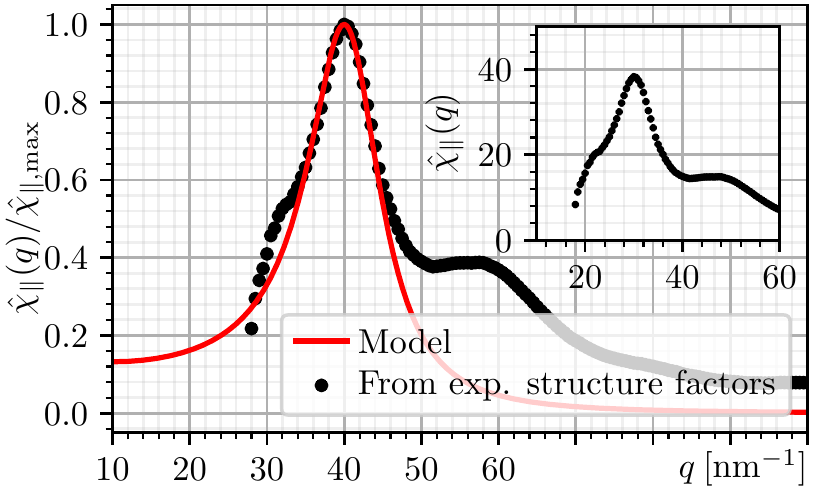} % AR : 0.6
	\caption{Dielectric susceptibility of bulk water.
	Black dots are recovered from inelastic neutron scattering data for oxygen-oxygen, hydrogen-hydrogen, and oxygen-hydrogen structure factors.\cite{bopp1996,soper1994}
	Red curve has been computed from Eq.~(\ref{chiparallel}) with $K=1/70$, $K_l = \SI{-2.01 E-3}{nm^{-2}}$, $\beta = \SI{1.12 E-6}{nm^{-4}}$.
	The inset shows the susceptibility which has not been normalized to 1.}
	\label{fig:1}
\end{figure}

% ---------------------------------------------------------------------------------------------- %
{\it Bulk water - }
The dielectric properties of bulk water are encoded in the two-points susceptibility tensor $\chi_{\alpha,\beta}(\vec{r}-\vec{r'})=\delta_{\alpha,\beta}(\vec{r}-\vec{r'})-\epsilon^{-1}_{\alpha,\beta}(\vec{r}-\vec{r'})$. 
This nonlocal kernel can be expressed through the correlations of the polarization ${\vec m}$  using the classical approximation for the fluctuation-dissipation theorem\footnote{There is a more general formulation taking into account quantum corrections\cite{bopp1998} that is not considered here for simplicity.},
\begin{equation}
\chi_{\alpha,\beta}(\vec{r}-\vec{r'})=\frac{\langle m_\alpha(\vec{r}) m_\beta(\vec{r'}) \rangle}{\epsilon_0 k_BT}.
\end{equation}
% Count : +16
The correlations $\langle m_\alpha(\vec{r}) m_\beta(\vec{r'}) \rangle$ can be written in terms of the experimentally measured partial HH, OH, OO structure factors of water\cite{soper1994} under the assumption of simple point charges localized at the atoms of molecules.\cite{bopp1996} 
The $q-$dependence of longitudinal part of the susceptibility in the Fourier space $\hat{\chi}_{\parallel}(q)$ illustrates the nonlocal nature of dielectric properties water (see Fig.~\ref{fig:1}). 
The main peak of  $\hat{\chi}_{\parallel}(q)$ (centered at $q=\SI{30}{nm^-1}$) exceeds 1, corresponding to a range of wavelengths associated with a negative permittivity $\epsilon_\parallel(q)=1/(1-\chi_\parallel(q))$.
This overscreening zone is a consequence of the H-bonding network in water\cite{bopp1996}.\par

To model these properties, we follow a Landau-Ginzburg (LG) approach which proved its value in the study of critical surface phenomena.\cite{lipowsky1983} 
We choose the following form of the second item in Eq.(\ref{elnonloc}),
\begin{equation}
	\label{elasticenergy}
	\mathcal{U}_m[\vec{m}] =
	\int{\frac{d \vec{r}}{2 \epsilon_0} \left( K \vec{m}^2 + 
	K_l \left(\vec{\nabla}\vec{m}\right)^2 +
	\beta \left(\vec{\nabla}\left(\vec{\nabla}\vec{m}\right)\right)^2\right)},
\end{equation} 
% Count : +32
which includes terms up to second spacial derivative of the field and leads to the longitudinal susceptibility, 
\begin{equation}
	\label{chiparallel}
	\hat{\chi}_{\parallel}(q)=\frac{1}{1 + K + K_l q^2 + \beta q^4}.
\end{equation}
% Count : +16
For derivation and discussion see the supporting material (SM).
The model parameters ($K$, $K_l$ and $\beta$) are chosen to capture: (i) the permittivity of bulk water at $q=0$, (ii) the position of the first peak and (iii) its width at half height of the simulated or experimentally recovered $\chi_\parallel(q)$.
The theoretical susceptibility is plotted in red in Fig.~\ref{fig:1}.

Its poles define a decay length $\lambda_d$ and a period $\lambda_o$,
\begin{equation}
	\lambda_d =\frac{2\sqrt{\beta}}{\sqrt{2\sqrt{(\beta (1+K)}+ K_l}}, \: \lambda_o = \frac{4 \pi \sqrt{\beta}}{ \sqrt{2\sqrt{\beta (1+K)} - K_l }},
\end{equation}
% Count : +32
characterizing the polarization correlations in bulk. They are equal to  $\lambda_d= \SI{2.1}{\angstrom}$ and $ \lambda_o = \SI{2.1}{\angstrom}$ for the chosen parametrization.  \par

% Count : + 110

\begin{figure} 
	\includegraphics{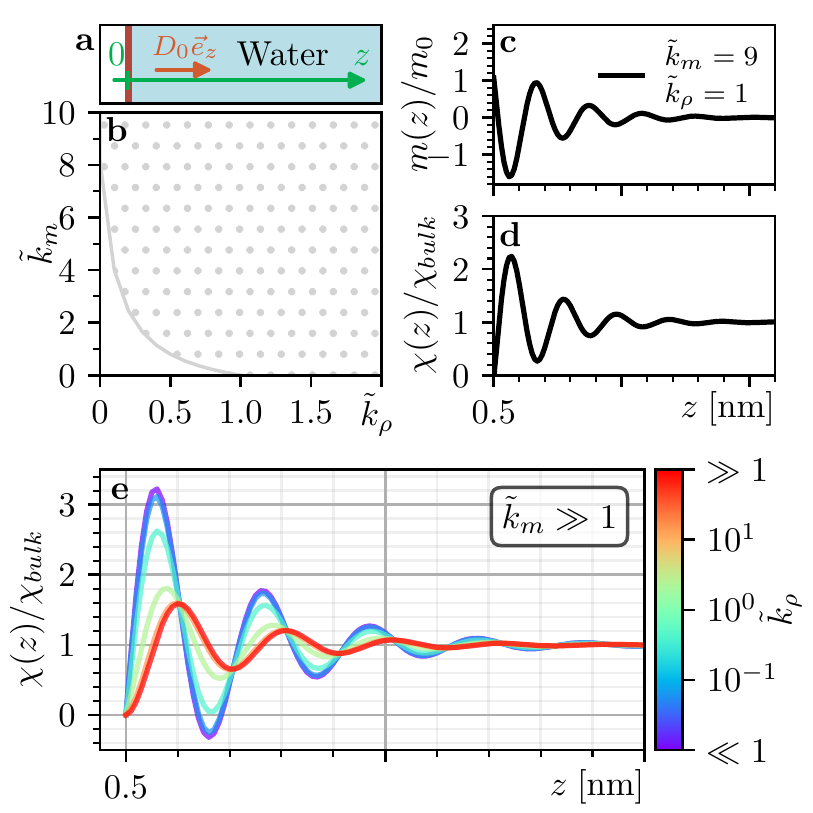} %AR : 1
	\caption{Dielectric properties of water in the vicinity of a surface.
	{\bf a.} Sketch of the system.
	{\bf b.} Diagram presenting the zone of finite minimum (dotted zone) as a function of $\tilde{k}_m$ and $\tilde{k}_\rho$.
	Profile of the polarization $m(z)$ \textbf{(c)} and the normalized susceptibility $\chi(z)$ \textbf{(d)} computed for ($\tilde{k}_m=9,\tilde{k}_\rho=1$) and ($m_0=\SI{-10}{V/nm},\rho_0/q_o=\SI{-10}{V/nm}$).
	{\bf e.} Susceptibility normalized to the bulk susceptibility computed from Eq.~(\ref{chikmlt1}) with different values of $\tilde{k}_\rho$.}
	\label{fig:2}
\end{figure}

% ---------------------------------------------------------------------------------------------- %
{\it Theoretical model for interfacial water - }
We consider water delimited by a planar interface infinite in the $xy$ plane and located at $z$=0 (See Fig.~\ref{fig:2}a).
A static homogeneous external field $\vec{D}_0 = D_0 \vec{u}_z$ is applied in the z-direction.
According to the symmetry of the problem, this field excites exclusively the longitudinal polarization that depends on $z$: $\vec{m}(\vec{r}) = m(z) \vec{u}_z$. 
We write the energy of the system per unit area $U[m,D_0]=U_{\mathrm{bulk}}+U_s$, the sum of the bulk-determined term, $U_{\mathrm{bulk}}$, derived from (Eqs.~\ref{elnonloc},\ref{elasticenergy}), and a surface term $U_s$ as
\begin{eqnarray}
	\label{Uw}
	U_{\mathrm{bulk}}&=&\int^{\infty}_{z=0}{\frac{dz}{2 \epsilon_0}\left[ \left( D_0 - m\right)^2+
K m^2+K_l \dot{m}^2+\beta \ddot{m}^2\right]} \nonumber\\
	U_s&=&\frac{k_m}{2}\left(m(0)-m_0)\right)^2+\frac{k_\rho}{2}\left(\rho(0)-\rho_0)\right)^2
\end{eqnarray}
% Count : +32
where the upper dot stands for the spatial derivation along $z$.
%the sum of the bulk-determined term, $U_{\mathrm{bulk}}$, and a surface term $U_s$.
In the spirit of the LG development used to express the kernel $\mathcal{K}$ (Eq.~(\ref{elnonloc})), $U_s$ is written as an expansion of elastic energies\cite{lipowsky1983,ajdari1992} depending on the polarization field and its derivative $\dot{m}(z)$, equal to minus the bound charge, $\rho(z)$.\cite{jackson1975} 
The major contribution promotes a surface polarization $m_0$ and the corrective second term favors a water charge density $\rho_0$ at the interface.
The stiffnesses $k_m$ and $k_\rho$ quantify the strength of the boundary conditions.
In the strong interaction limit $(k_m,k_\rho)\rightarrow \infty$, the surface fixes both polarization and charge density at interface. \par

The partition function of the system,
$\mathcal{Z}[D_0] = \int{D[m_z] \exp{\left[-\left( U_{\mathrm{bulk}}[m,D_0] + U_s\right)/k_BT \right]}},$
can be split in the form
\begin{widetext}
\begin{equation}
	\mathcal{Z} [D_0] = 
	\int d \bar{m} d \bar{\rho} \exp{\left[\frac{1}{k_BT}
		\left(\frac{k_m}{2} (\bar{m}-m_0)^2 + \frac{k_\rho}{2} (\bar{\rho}-\rho_0)^2\right)\right]}
	\int^{\substack{m(z\rightarrow\infty)=0 \\ \dot{m}(z\rightarrow\infty)=0}}_{\substack{m(0)=\bar{m} \\ \dot{m}(0)=-\bar{\rho}}}{D[m] \exp{\left[-\frac{1}{k_BT}U_{\mathrm{bulk}}[m,D_0] \right]}}.
\end{equation}
\end{widetext}
% Count : +32
This includes a partition of the fields $m(z)$ satisfying the boundary conditions
%$\left(m(0)=\bar{m}, \dot{m}(0)=-\bar{\rho}\right)$
 (right integral),
% $(m(z\rightarrow\infty)=0, \dot{m}(z\rightarrow\infty)=0)$, 
then a sampling of the $z=0$ boundary conditions $(\bar{m}, \bar{\rho})$ (left integral).
We find the mean field solution, $m_1(z)$, by first minimizing $U_{\mathrm{bulk}}[m, D_0]$ with respect to $m(z)$ leading to
\begin{eqnarray}
	\label{edplanar}
	(1+K) m(z) - K_l m^{(2)}(z) + \beta m^{(4)}(z) = D_0, 
\end{eqnarray}
% Count : +16
with $m(0)=\bar{m}$, $\dot{m}(0)=-\bar{\rho}$, $m(z\rightarrow\infty)=0$, $\dot{m}(z\rightarrow\infty)=0 $.
The solution of which is
\begin{eqnarray}
	\label{mtest}
&	&m_1(z)=\frac{D_0}{1+K} \left(1-e^{-\frac{z}{\lambda_d}}\left(\cos(q_oz)+\frac{q_d}{q_o}\sin(q_oz)\right)\right)\nonumber\\&
	+&e^{-\frac{z}{\lambda_d}}\left(\bar{m}\left(\cos(q_oz)+\frac{q_d}{q_o}\sin(q_oz)\right)
	-\frac{\bar{\rho}}{q_0}\sin(q_oz)\right).
\end{eqnarray}
% Count : +32
with $q_o=2 \pi/\lambda_o$ and $q_d=1/\lambda_d$, the wavenumbers of the bulk correlations.
Second, we extremalize the total energy of the system, $U=U_{\mathrm{bulk}}+U_s$, with respect to $(\bar{m},\bar{\rho})$ obtained by injecting $m_1(z)$ in Eqs.~(\ref{Uw}) and performing the integral over $z$ (see SM). 
The nature of the extremum depends on the dimensionless stiffness constants $(\tilde{k}_m,\tilde{k}_\rho)$, given in the SM. 
$U(\bar{m},\bar{\rho})$ admits a minimum for $\tilde{k}_m$ and $\tilde{k}_\rho$ belonging to the pointed zone represented in Fig.~\ref{fig:2}b, to which we restrict our study in the following. The 
mean field polarization $m_2(z)$ is given by Eq.~(\ref{mtest}) for $\bar{m}=m_s,\bar{\rho}=\rho_s$,
the boundary conditions minimizing $U(\bar{m},\bar{\rho})$. Their expressions are given in SM.\par 
To study the dielectric properties of interfacial water, we introduce the real space susceptibility, $\chi(z)=d m_2(z)/d D_0$ derived from Eq.~(\ref{mtest}). It quantifies the response to a homogeneous external field $D_0$ and is constant and equal to $\chi_b=\hat{\chi}(0)$ for bulk water.\par

Fig.~\ref{fig:2}c, d show typical mean field polarization $m_2(z)$ and susceptibility $\chi(z)$ in the interfacial water.
We observe a nonvanishing polarization  and a nonconstant $\chi(z)$ that are oscillating functions of period $\lambda_o$ in an exponentially decaying envelope of range $\lambda_d$.
The surface induces a layering of the fluid that extends over about $\SI{1}{nm}$, a lengthscale consistent with many previous simulations of interfacial water\cite{bonthuis2012,schlaich2016}. 
The susceptibility shows alternation of underresponding ($\chi(z) \ll \chi_b$) and overresponding layers ($\chi(z)\gg \chi_b$), typical for overscreening effect.
Whereas the amplitude of $m_2(z)$ is a non-trivial function of the bulk properties, the four parameters of the surface interaction and $D_0$,  the amplitude of $\chi(z)$ does not depend on ($m_0,\rho_0,D_0$). The interface affects the dielectric properties of water only through the stiffnesses $(\tilde{k}_m, \tilde{k}_\rho)$.\par
 
We first study the case of vanishing $\tilde{k}_\rho$ for which $\chi(z)_{\tilde{k}_\rho=0}$ can be written as
\begin{equation}
	\label{chikrho0}
	\frac{\chi(z)_{\tilde{k}_\rho=0}}{\chi_b} = 1 + \frac{\tilde{k}_m e^{-q_dz}}{1-\tilde{k}_m}\left(\cos(q_oz) + \frac{q_d^2 - q_o^2}{2q_d q_o}\sin(q_oz)\right).
\end{equation}% Count : +32
The amplitude of $\chi(z)_{\tilde{k}_\rho=0}$ decreases with $\tilde{k}_m$ and tends to a finite value for $\tilde{k}_m\gg1$. This case is represented in Fig.~\ref{fig:2}e (blue curve).\footnote{Note that the diverging value $\tilde{k}_m=1$ corresponds to the boundary of the stability region of the phase space parameter (see Fig.~\ref{fig:2}b)}
Then we consider the corrective effect of $\tilde{k}_\rho$ in the limit of a large $\tilde{k}_m$ by studying
\begin{equation}
\label{chikmlt1}
\frac{\chi(z)_{ \tilde{k}_m\gg 1}}{\chi_b} = 
\frac{\chi(z)_{ \tilde{k}_m\gg 1, \tilde{k}_\rho=0}}{\chi_b} - 
\frac{q_d^2 +q_o^2}{2 q_d q_o} \frac{\tilde{k}_\rho e^{-q_dz}}{1+\tilde{k}_\rho} \sin(q_oz).
\end{equation} 
% Count : +32
An increasing $\tilde{k}_\rho$ induces a dephasing  and an amplitude decrease up to a factor 2 of $\chi(z)$   (See ~\ref{fig:2}e).
The behavior of $\chi(z)$ as a function of $(\tilde{k}_m,\tilde{k}_\rho)$ illustrates that different surfaces, having stronger or weaker influence on polarization and partial charge, induce different dielectric properties of interfacial water.\par
 
% Count : +170

% ---------------------------------------------------------------------------------------------- %
{\it Comparison with MD simulations - }
We performed MD simulations of pure water confined in a slab geometry using the GROMACS MD simulation package.\cite{lindahl2018}
Water molecules are described with the SPC/E model and
the walls are made up of atoms of frozen positions. We considered graphene and hBN surfaces ( details in the SM).\par

\begin{figure}
	\includegraphics{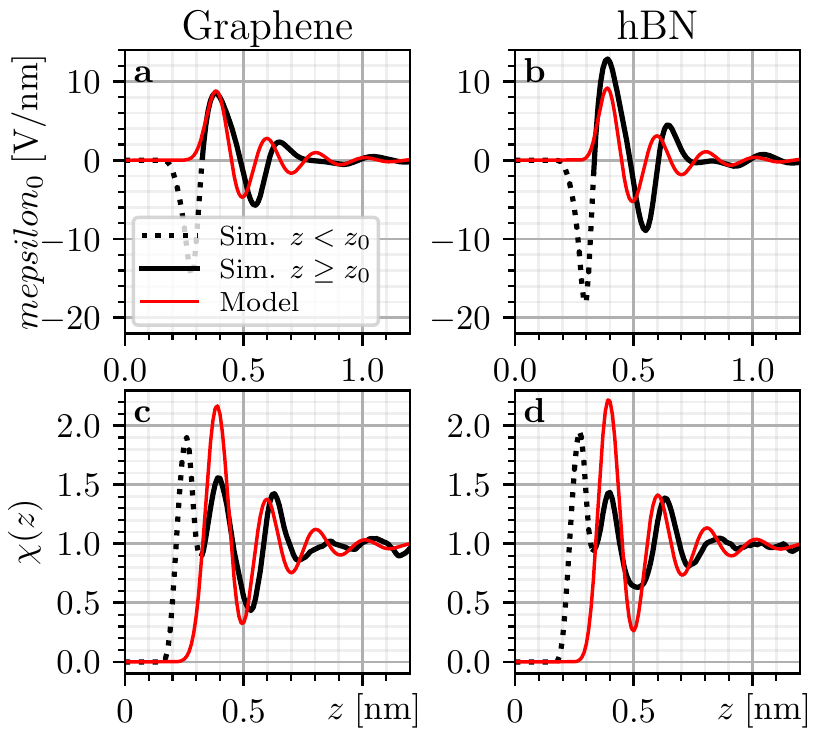} %AR : 1.2
	\caption{Comparison between model (in red) and MD simulations (in black) for a graphene layer (left panels) and a hBN layer (right panels).
	Top (respectively bottom) panels show the polarization (respectively the susceptibility).
	Simulation curves for $z \le z_0$ are represented with dotted lines. }
	\label{fig:3}
\end{figure}

We analyze the polarization, $m_{\rm{MD}}(z)=-\int^z_0 dz \rho_{\rm{MD}}(z) dz$,  with $\rho_{\rm{MD}}$ the charge density of water, and the susceptibility  $\chi_{\rm{MD}}(z) = (m_{\rm{MD}}(z,D_0+\delta D_0)-m_{{\rm{MD}}}(z,D_0))/\delta D_0$ \cite{bonthuis2012} with $\delta D_0 = 0.5V/nm$, in the vicinity of the surfaces. 
The profiles are similar for both surfaces (Fig.~\ref{fig:3}): first, a vacuum layer ($m_{\rm{MD}}(z)=0$, $\chi_{\rm{MD}}(z)=0$) between the surface and the liquid, due to the repulsive part of the surface-fluid  Lennard-Jones (LJ) interaction, then decaying oscillations over about $\SI{1}{nm}$ before reaching the bulk value.
The theoretical decay $\lambda_d$ and the period $\lambda_o$ are in very good agreement with the simulated ones (see SM). This validates the derivation of the characteristic lengths of interfacial water from the bulk dielectric susceptibility, $\hat{\chi}(q)$.\par
 
In MD simulations, the position of the interfaces is not as clear-cut as in theory due to thermal capillary fluctuations and the non-infinitly sharp repulsion of the surface-fluid LJ interaction.\cite{yang2019}
This is taken into account by applying a smearing to the theoretical predictions,
\begin{equation}
	\label{smearedfunctions}
	\tilde{f}(z) = \left(G \ast (\theta f)\right)(z+z_0),
	\:
	G(z) = \frac{e^{-z^2/2\eta^2}}{\eta\sqrt{2\pi}}, 
\end{equation}
% Count : +16
with $\theta$ being the Heaviside function and $f$ standing for $m$ or $\chi$. The position $z_0$ and the standard deviation $\eta$ are determined for each surface by fitting the first oxygen density peak with a Gaussian $G(z)$ which position and width define $z_0$
 and $\eta$ (see SM for details).
The hBN surface is characterized by a deeper LJ potential and consequently a smaller dispersion $\eta$ than the graphene.
Correspondly, $m_{\rm MD}(z)$ amplitude is smaller in interfacial water for graphene than for hBN Figs.~\ref{fig:3}a-b.\par
 
\par
% Count : +200
We validate the theoretical model in three steps.
First, we adjust the simulated susceptibilities with $\tilde{\chi}(z)$ defined in Eq.~(\ref{smearedfunctions}).
If we choose ($\tilde{k}_m\gg1$, $\tilde{k}_\rho=0$) for graphene and ($\tilde{k}_m\gg1$, $\tilde{k}_\rho=0.2$) for hBN, we obtain a good agreement between the calculated and the simulated 
value of the susceptibilities as shown in figures~\ref{fig:3}c-d.
Next, we fit the simulated polarization for graphene surface with $\tilde{m}(z)$ by fixing $m_0$, the single left unknown parameter for graphene as $\tilde{k}_\rho=0$. Finally, we fit the simulated polarization for a hBN surface.
Taking the surface polarization $m_0$ previously determined in the case of graphene, we fix $\rho_0$.
The comparison between theoretical and simulated polarization are presented in Figs.~\ref{fig:3}a-b. 
The dotted part of the simulated curves correspond to the vacuum gap and the contribution of hydrogen located in $z<z_0$. The theoretical model describes this zone as a vacuum gap (see Eq. (\ref{smearedfunctions})).\par
 
Graphene and hBN surfaces are parametrized by $\tilde{k}_m \gg 1$, thus both surfaces freeze the interfacial polarization to  $m(z_0)=m_0$ which does respond to $D_0$.
 At the microscopic scale, this result can be interpreted as the effect of the vacuum gap on the organization in the first layer of water which optimizes the number of H-bonds.\cite{varghese2019}
Most likely, $\tilde{k}_m$ is very large for a wide variety of surfaces, both hydrophobic and hydrophilic, as they impose a layout in the first hydration layer.\cite{bonthuis2012, besford2020} 
For a non-vanishing corrective term $\tilde{k}_\rho$, the surface has an effect on the interfacial charge, $\rho_s(z_0)$, and its variation under $D_0$.
We investigate the microscopic origin of this effect by performing MD simulations for artificial surfaces associated with hybrid properties between graphene and hBN surfaces (see SM).  We find out that
it is induced by a large mean depth of the LJ minimum. A non-vanishing $\tilde{k}_\rho$ is related with important variations of the interaction energy between the surface and a water molecule in the $(xy)$ plane for $z=z_0$ that constrains the position of water molecules in this plane.  \par

\begin{figure}
	\includegraphics{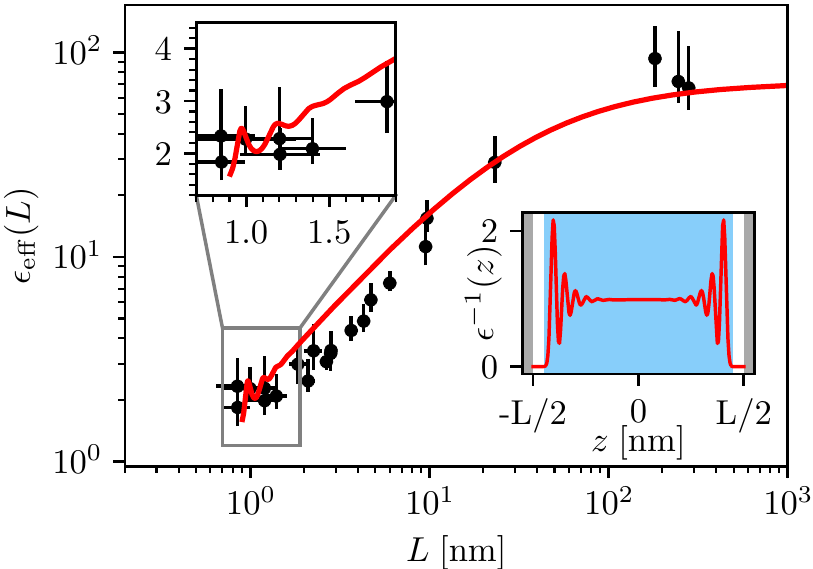}
	\caption{Effective dielectric permittivity $\epsilon_{{\rm eff}}$ of water nanoconfined in a channel of width $L$.
	Comparison between experimental measurements reproduced from \cite{fumagalli2018} and theoretical model. }
	\label{fig:4}
\end{figure}\par
% Count : +140

{\it Nanoconfined water - } We use now this theoretical model to derive the dielectric properties of a confined water layer.
The experimental measurements report an effective dielectric permittivity up to $\epsilon_{{\rm eff}}=2$ for a channel of about $\SI{1}{nm}$.\cite{fumagalli2018} (reported on Fig.~\ref{fig:4}).
The authors suggest the existence in the channel of three water layers of homogeneous dielectric properties: two interfacial layers ($\epsilon=2.1$, thickness: $\SI{0.7}{nm}$) and a layer of bulk water ($\epsilon=78$). 
We compute the effective permittivity $\epsilon_{{\rm eff}} = L/\int_0^L(1-\chi(z))dz$ as a function of $L$ for two graphene surfaces.  Our model can be seen as two vacuum gaps and an inhomogeneous water layer.
This inhomogeneity is not implemented {\it ad hoc} but is the signature of the nonlocal dielectric properties of water, revealed by the boundary conditions.
The results are presented in Fig.~\ref{fig:4}.
The model reproduces the experimental measures and catches in particular a non-homogenous behavior of the permittivity as a function of $L$ for small $L$ as shown in the insert that cannot be described by a three homogeneous layer model.\cite{loche2020,fumagalli2018,zhang2018}\par

% ---------------------------------------------------------------------------------------------- %
{\it Conclusion - } Nanoconfined water is a non-homogeneous dielectric material which properties differ dramatically from the bulk.
Water-surface interaction and the bulk properties of the fluid combine to induce specific dielectric profiles.
The complexity of this system is captured by a minimal phenomenological Hamiltonian depending on the polarization field and composed of (i) a LG forth order development for a bulk-determined term and (ii) a harmonic surface-water term.
We show that the dielectric susceptibility of interfacial water may be strongly affected by the coarse-grained parameters ($\tilde{k}_m$, $\tilde{k}_\rho$, $\eta$) characterizing local surface-water interaction.
It gives a framework to compare graphene and hBN that could be predictive for other surfaces and also to derive the dielectric properties confined water in other geometries, such as nanotubes.\cite{loche2019}\par

\begin{acknowledgments}
This work was supported by Sorbonne Sciences under grant Emergences-193256. HB and GM thank B. Delamotte for fruitful discussions.
\end{acknowledgments}

% Bibliographies
\bibliographystyle{apsrev4-2}
\bibliography{references}

% Word count
% \verbatiminput{\jobname.wordcount.tex}
% https://journals.aps.org/authors/length-guide
% Total Word Count = Text + Displayed Math + Figures + Tables
% Displayed Math	The word equivalent for displayed math is 16 words per row for single-column equations. Two-column equations count as 32 words per row.
% Figures : +620
% Equations : +320
\end{document}

\title{Supplementary Materials for\\The nonlocal dielectric response of water in nanoconfinement}

\author{G. Monet} 
\affiliation{Sorbonne Universit{\'e}, CNRS, Laboratoire de Physique Th{\'e}orique de la Mati{\`e}re Condens{\'e}e (LPTMC, UMR 7600), F-75005 Paris, France}
\author{F. Bresme}
\affiliation{Department of Chemistry, Molecular Sciences Research Hub ant Thomas Young Center for Theory and Simulations of Materials, Imperial College London,  W12 0BZ 2AZ London, United Kingdom}
\author{A. Kornyshev}
\affiliation{Department of Chemistry, Molecular Sciences Research Hub ant Thomas Young Center for Theory and Simulations of Materials, Imperial College London,  W12 0BZ 2AZ London, United Kingdom}
\author{H. Berthoumieux} 
\affiliation{Sorbonne Universit{\'e}, CNRS, Laboratoire de Physique Th{\'e}orique de la Mati{\`e}re Condens{\'e}e (LPTMC, UMR 7600), F-75005 Paris, France}

\maketitle

\tableofcontents

\section{The theoretical model}
\subsection{Bulk water}
The functional energy $\mathcal{U}_{\textrm{bulk}}$ of the dielectric medium modeling bulk water is written as
\begin{align}
	\label{elnonloc}
		\mathcal{U}_{\textrm{bulk}}[\vec{m},\vec{D}_0] &= 
        \frac{1}{2 \epsilon_0} \int{d \vec{r} \left( \vec{D}_0(\vec{r}) - \vec{m}(\vec{r}) \right)^2} 
        +\frac{1}{2 \epsilon_0}\int{d \vec{r} d \vec{r'} m^\alpha(\vec{r}) \mathcal{K}_{\alpha,\beta}(\vec{r},\vec{r'}) m^\beta(\vec{r'})}\\
        &= 
        \begin{aligned}[t]
            &\frac{1}{2 \epsilon_0} \int{d \vec{r} \left( \vec{D}_0^2(\vec{r}) - 2 \vec{D}_0(\vec{r}) \cdot \vec{m}(\vec{r}) \right)} \\
            &+\frac{1}{2 \epsilon_0} \int{d \vec{r} d \vec{r'} m^\alpha(\vec{r}) \left(\delta_{\alpha, \beta}(\vec{r} - \vec{r'}) + \mathcal{K}_{\alpha,\beta}(\vec{r},\vec{r'}) \right) m^\beta(\vec{r'})},
        \end{aligned}
\end{align} 
where we use the Einstein summation convention over repeated indices.\par

The dielectric material is homogeneous and isotropic so $\mathcal{K}_{\alpha,\beta}(\vec{r},\vec{r'}) = \mathcal{K}_{\alpha,\beta}(\left| \vec{r'} - \vec{r} \right|)$ and we can write the non-local contribution of the functional energy (Eq.~(\ref{elnonloc})) as function of the Fourier transform of the polarization field $\hat{m}(\vec{q}) = \frac{1}{(2\pi)^{3/2}} \int{d\vec{r}\;\vec{m}(\vec{r}) e^{-i \vec{q} \vec{r}}}$ and introduce the dielectric susceptibility tensor $\hat{\mathcal{\chi}}_{\alpha, \beta}^{-1}(\vec{q})$ :
\begin{equation}
    \int{d \vec{r} d \vec{r'} m^\alpha(\vec{r}) \left(\delta_{\alpha, \beta}(\vec{r} - \vec{r'}) + \mathcal{K}_{\alpha,\beta}(\vec{r},\vec{r'}) \right) m^\beta(\vec{r'})} = 
    \frac{1}{(2\pi)^{3/2}} 
    \int{d\vec{q} \hat{m}^\alpha(\vec{q}) \hat{\mathcal{\chi}}_{\alpha, \beta}^{-1}(\vec{q}) \hat{m}^\beta(-\vec{q}) }.
\end{equation}
We then write the functional of bulk energy $\mathcal{U}_{\mathrm{bulk}}$ in Fourier space:
\begin{equation}
    \label{eqn:U_bulk_TF}
    \mathcal{U}_{\textrm{bulk}}[\vec{\hat{m}},\vec{\hat{D}}_0] = \frac{1}{2\epsilon_0} \frac{1}{(2\pi)^{3/2}} 
    \int{d\vec{q}
        \left[
            \vec{\hat{D}}_0(\vec{q}) \vec{\hat{D}}_0(-\vec{q})  -2  \vec{\hat{m}}(-\vec{q}) \vec{\hat{D}}_0(\vec{q}) +
            \hat{m}^\alpha(\vec{q}) \hat{\mathcal{\chi}}_{\alpha, \beta}^{-1}(\vec{q}) \hat{m}^\beta(-\vec{q})
        \right]
    }.
\end{equation}
The mean field polarization, $\vec{\hat{m}}_{\rm MF}$, is defined as the minimum of the functional energy:
\begin{equation}
    \frac{\delta \mathcal{U}_{\textrm{bulk}}}{\delta \vec{\hat{m}}}[\vec{\hat{m}}_{\rm MF}]=0.
\end{equation}
By performing the functional derivative of the equation~\ref{eqn:U_bulk_TF}, we show:
\begin{equation}
        \hat{\vec{D}}_{0}(\vec{q})  =  \hat{\mathcal{\chi}}^{-1}(\vec{q}) \hat{\vec{m}}_{\rm MF}(\vec{q}) 
        \Leftrightarrow 
        \hat{\vec{m}}_{\rm MF}(\vec{q}) = \hat{\mathcal{\chi}}(\vec{q}) \hat{\vec{D}}_{0}(\vec{q}) 
\end{equation}\par
Since the dielectric material is homogeneous and isotropic, the dielectric susceptibility tensor $\hat{\mathcal{\chi}}^{-1}$ has two components :
\begin{equation}
    \hat{\mathcal{\chi}}_{\alpha,\beta}^{-1}(\vec{q}) = 
    \hat{\chi}_{\parallel}^{-1}(q) \frac{q_\alpha q_\beta}{q^2} +
    \hat{\chi}_{\perp}^{-1}(q) \left( \delta_{\alpha, \beta} - \frac{q_\alpha q_\beta}{q^2}\right),
\end{equation}
where $\hat{\chi}_{\parallel}^{-1}$ is the longitudinal component and $\hat{\chi}_{\perp}^{-1}$, the transverse one.
It can be seen that developing the longitudinal component to the fourth order,
\begin{eqnarray}
    \hat{\chi}_{\parallel}^{-1}(q)  &=& 1 + K +K_l q^2 + \beta q^4 \\
    \hat{\chi}_{\perp}^{-1}(q)  &=& K,
\end{eqnarray}
is equivalent to the Landau Ginzburg approach used in the main article:
\begin{equation}
    \label{eqn:elnonloc4}
    \mathcal{U}_{\textrm{bulk}}[\vec{m},\vec{D}_0] =
    \frac{1}{2 \epsilon_0} \int{d \vec{r} \left( \vec{D}_0 - \vec{m}(\vec{r}) \right)^2} 
    +\frac{1}{2 \epsilon_0}\int{d \vec{r} \left( K \vec{m}^2 + 
    K_l \left(\vec{\nabla}\cdot\vec{m}\right)^2 +
    \beta \left(\vec{\nabla} \left(\vec{\nabla} \cdot \vec{m}\right)\right)^2\right)}.
\end{equation}

\subsection{Polarization  for one-wall geometry system}
\begin{figure}[H]
    \centering 
    \includegraphics{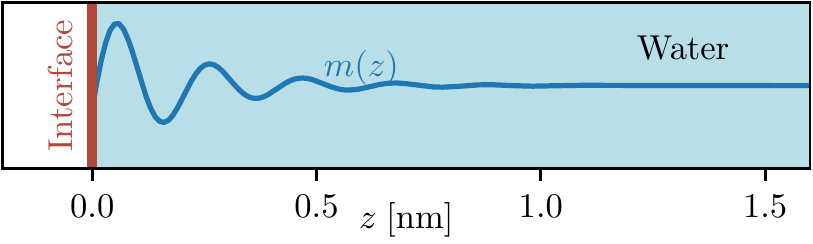}
    \caption{System with water close to a wall. The polarization field is schematically shown with the blue curve (computed from Eq.~\ref{mtest}).}
    \label{SI-fig:scheme_wall}
\end{figure}\par

As stated in the main text, there is a polarization field $m_1(z)$ that minimizes the bulk part of the functional of the energy $U_{\mathrm{bulk}}$ defined in equation~8 : 
\begin{eqnarray}
	\label{mtest}
	m_1(z)&=&\frac{D_0}{1+K} \left(1-e^{-\frac{z}{\lambda_d}}\left(\cos(q_oz)+\frac{q_d}{q_o}\sin(q_oz)\right)\right)\nonumber\\&
	+ &\bar{m}e^{-\frac{z}{\lambda_d}}\left(\cos(q_oz)+\frac{q_d}{q_o}\sin(q_oz)\right)
	-\lambda_o\bar{\rho}e^{-\frac{z}{\lambda_d}}\sin(q_oz).
\end{eqnarray}
with $m(0)=\bar{m}$ and $\dot{m}(0) = -\bar{\rho}$. 
We introduce this expression of polarization into the total functional of energy
\begin{equation}
    \begin{split}
        U=U_{\mathrm{bulk}}+U_s =	\frac{1}{2 \epsilon_0}\int^{+\infty}_{z=0}{dz \left( D_0 - m(z) \right)^2}+\frac{1}{2 \epsilon_0}
            \int^{+\infty}_{z=0}{dz K m(z)^2+K_l (\dot{m}(z))^2+\beta (\ddot{m}(z))^2} \\
            +\frac{k_m}{2}(\bar{m}-m_0)^2+\frac{k_\rho}{2}(\bar{\rho}-\rho_0)^2,
    \end{split}
\end{equation}
where dots stand for $d/dz$.
Performing the integral over $z$ in $U_{\mathrm{bulk}}$, we express  $U$ as a function of the boundary constants  ($\bar{m},\bar{\rho}$):
\begin{equation}\begin{split}
U(\bar{m},\bar{\rho})=  
& \frac{k_m}{2}(\bar{m}-m_0)^2 +\frac{k_\rho}{2}(\bar{\rho}-\rho_0)^2 +\frac{\qd D_0}{(\qd^2 + \qo^2) (1+K) \epsilon_0} \\
& + \frac{D_0}{(\qd^2 + \qo^2) \epsilon_0} \bar{\rho} +
\frac{\qd (1+K)}{(\qd^2 + \qo^2)^2 \epsilon_0} \bar{\rho}^2\\  &
- \frac{2 \qd D_0}{(\qd^2 + \qo^2) \epsilon_0} \bar{m} +
\frac{\qd (1+K)}{(\qd^2 + \qo^2) \epsilon_0}\bar{m}^2 -
\frac{1+K}{(\qd^2 + \qo^2) \epsilon_0} \bar{\rho} \bar{m}.
\end{split}\end{equation}
We find the stationary point of the energy $U(\bar{m},\bar{\rho})$  by cancelling its gradient:
\begin{equation}
    \left\{
        \begin{aligned}
        \frac{\partial U}{\partial \bar{m}} (m_s, \rho_s)=0\\ 
        \frac{\partial U}{\partial \bar{\rho}} (m_s, \rho_s)=0
        \end{aligned}
    \right.
    \Rightarrow 
    \left\{
        \begin{aligned}
            -\frac{\left(q_d^2+q_o^2\right)}{2q_d} \left(m_s - \frac{D_0}{1+K}\right)
            +\rho_s +\tilde{k}_\rho (\rho_s - \rho_0) = 0\\
            2q_d \left(m_s - \frac{D_0}{1+K}\right) 
            -\rho_s - \frac{3q_d^2 - q_o^2}{2q_d} \tilde{k}_m (m_s - m_0) = 0
        \end{aligned}
    \right.
\end{equation}
where $\tilde{k}_m = k_m/k_{m,\rm{bulk}}$ and $\tilde{k}_\rho = k_\rho/k_{\rho,\rm{bulk}}$ are dimensionless stiffness constants with 
\begin{eqnarray}
    k_{m,\rm{bulk}} &=& - \frac{\left(3 q_d^2 - q_o ^2\right) (1+K)}{2q_d \left(q_d^2 + q_o^2\right) \epsilon_0} \\
    k_{\rho,\rm{bulk}} &=& \frac{2q_d (1+K)}{\left(q_d^2 + q_o^2\right)^2 \epsilon_0}.
\end{eqnarray}

The expression of the stationary point $(m_s, \rho_s)$ is linear function of the external field $D_0$ and be written as
\begin{eqnarray}
    \rho_s(D_0) &=& a_\rho \frac{D_0}{1+K} + b_\rho \\
    m_s(D_0) &=& a_m \frac{D_0}{1+K} + b_m,
\end{eqnarray}
with 
\begin{eqnarray}
a_\rho &=& -\frac{\left(3q_d^2 - q_o^2\right) \left(q_d^2 + q_o^2\right) \tilde{k}_m}{2 q_d \left(\left(3 q_d^2 - q_o^2\right) \left(\tilde{k}_m \left(1 + \tilde{k}_\rho\right) - 1\right) -4q_d^2 \tilde{k}_\rho\right)} \\
a_m &=& -\frac{\left(3q_d^2 - q_o^2\right) + 4 q_d^2 \tilde{k}_\rho}{\left(3 q_d^2 - q_o^2\right) \left(\tilde{k}_m \left(1 + \tilde{k}_\rho\right) - 1\right) -4q_d^2 \tilde{k}_\rho}
\end{eqnarray}
and
\begin{eqnarray}
b_\rho &=& \frac{\left(3q_d^2 - q_o^2\right) \tilde{k}_m \left(\left(q_d^2 + q_o^2\right) m_0 + 2 q_d \rho_0 \tilde{k}_\rho\right) -8 q_d^3 \rho_0 \tilde{k}_\rho}{2 q_d \left(\left(3 q_d^2 - q_o^2\right) \left(\tilde{k}_m \left(1 + \tilde{k}_\rho\right) - 1\right) -4q_d^2 \tilde{k}_\rho\right)} \\
b_m &=& \frac{\left(3 q_d^2 - q_o^2\right) m_0 \tilde{k}_m \left(1 + \tilde{k}_\rho \right) - 2 q_d \rho_0 \tilde{k}_\rho}{\left(3 q_d^2 - q_o^2\right) \left(\tilde{k}_m \left(1 + \tilde{k}_\rho\right) - 1\right) -4q_d^2 \tilde{k}_\rho}.
\end{eqnarray}

\begin{figure}[H]
    \centering 
    \includegraphics{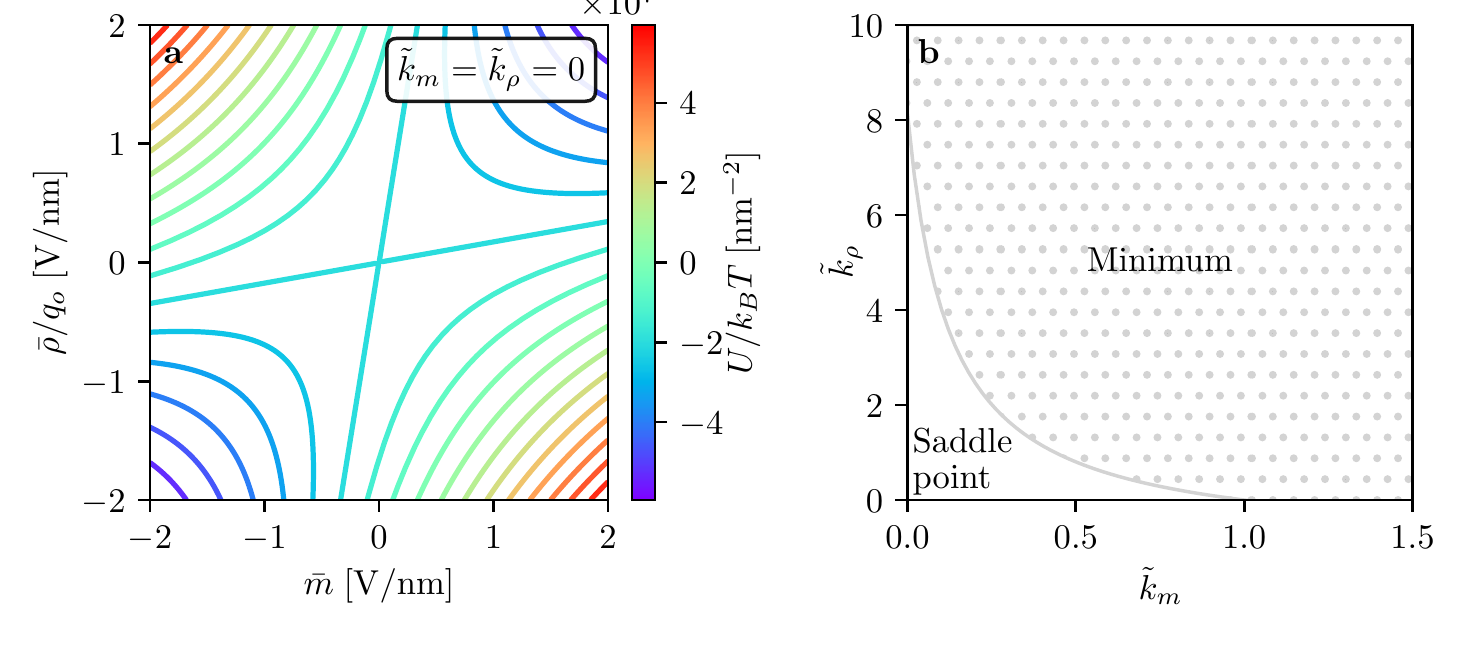}
    \caption{\textbf{a}. Function $U(\bar{m},\bar{\rho})$ for $(\tilde{k}_m=0,\tilde{k}_\rho=0)$ and a vanishing value of $D_0$.
    \textbf{b}.  The dotted area represents a zone of parameters $\tilde{k}_m$ and $\tilde{k}_\rho$ for which the function $U(\bar{m},\bar{\rho})$ possesses a minimum for finite value of $\bar{m}$ and $\bar{\rho}$.
    Outside this area the functional energy shows a saddle point as depicted in the left panel.}
    \label{SI-fig:instability}
\end{figure}

The stationary point $(m_s, \rho_s)$ is not always a minimum as evidenced by the determinant $  \mathrm{Det}(\mathbf{H} \; U_{w}) $ of the Hessian matrix,
\begin{eqnarray}
	\mathbf{H}=\left(\begin{array}{cc}
	\frac{\partial ^2 U_{\mathrm{bulk}}}{\partial \bar{m}\partial \bar{m}} &
	\frac{\partial ^2U_{\mathrm{bulk}}}{\partial \bar{\rho} \partial \bar{m}}\\\frac{\partial ^2 U_{\mathrm{bulk}}}{\partial \bar{\rho} \partial \bar{m} }& \frac{\partial ^2U_{\mathrm{bulk}}}{\partial \bar{\rho} \partial \bar{\rho}}
	\end{array}\right).
\end{eqnarray}
We find that 
\begin{equation}
    \label{eqn:hessianDet}
    \mathrm{Det}(\mathbf{H})(\tilde{k}_m, \tilde{k}_\rho) = \frac{(\qo^2 - 3\qd^2 )(1+K)^2}{\epsilon^2 \left(\qd^2 + \qo^2\right)^3 }
    \left(\tilde{k}_m \left(1+\tilde{k}_\rho\right) -\frac{4 q_d^2}{3q_d^2 - q_o^2} \tilde{k}_\rho 
    - 1\right)
\end{equation}
can be negative for given values of $(\tilde{k}_m, \tilde{k}_\rho)$.
Under these conditions, the functional energy $U(\bar{m},\bar{\rho})$ extremum is a saddle point.
This saddle point is actually shown in figure~\ref{SI-fig:instability}a  for a non-interaction surface $(\tilde{k}_m=0,\tilde{k}_\rho=0)$ and a vanishing value of $D_0$.
The polarization and the layering of the fluid by a non-interacting interface, {\it i.e. } by a pure geometrical constraint, is a consequence of the pronounced overscreening associated with specific wavelengths.
This non-physical limit could be avoided by adding a non-linear term in $m^4$ in Eq.~(\ref{eqn:elnonloc4}) that would ensure a saturation effect in the fluid.
In Figure~2a of the main article and Figure~\ref{SI-fig:instability}b, the gray curve separating the region for which the extremum of the energy functional $U(\bar{m},\bar{\rho})$ is a minimum (gray dotted area) and that for which this extremum is a saddle point (white area) is computed by cancelling equation~\ref{eqn:hessianDet}.

\par

\subsection{Water confined in between two surfaces : a slab geometry}
\begin{figure}[H]
    \centering 
    \includegraphics{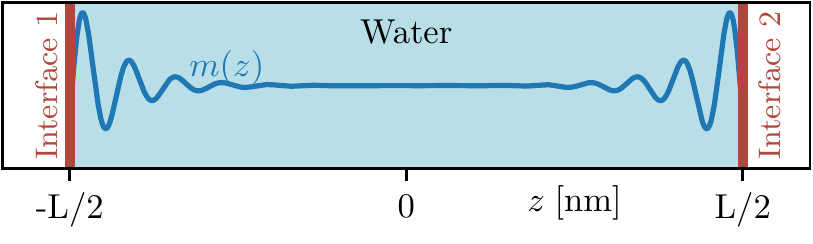}
    \caption{System with water confined between two walls separated by a length $L$.
    The blue curve is a schematic representation of the polarization highlighting the oddity of this function for vanishing $D_0$.}
    \label{SI-fig:scheme_slab}
\end{figure}\par

We now consider a system composed of two identical surfaces, located in $z=-L/2$ and $z=L/2$ as represented in Figure~\ref{SI-fig:scheme_slab}.
In this case, the functional energy writes is
\begin{equation}
    U = U_{\mathrm{bulk}}+ U_s,
\end{equation}
the sum of the electrostatic energy of water $U_{\mathrm{bulk}}$ and the interaction energy $U_s$ 
with 
\begin{eqnarray}
    U_{\mathrm{bulk}}[m] &=& \frac{1}{2 \epsilon_0} \int_{-L/2}^{L/2}{
        (m(z)-D_0)^2 dz
    }+ \frac{1}{2 \epsilon_0} \int_{-L/2}^{L/2}{
        \left( K m^2(z) + 
        K_l \dot{m}^2(z) +
        \beta \ddot{m}^2(z)\right) dz
    }\\
    U_s &=& 
    \frac{k_\rho}{2} \left(\bar{\rho}_- - \rho_0 \right)^2 + 
    \frac{k_\rho}{2} \left(\bar{\rho}_+ - \rho_0 \right)^2 +
    \frac{k_m}{2}      \left(\bar{m}_- - m_0 \right)^2 +
    \frac{k_m}{2}       \left(\bar{m}_+ + m_0 \right)^2
\end{eqnarray}
and
\begin{equation}
    \begin{aligned}
        \bar{\rho}_- = \rho(z=-L/2) &\qquad& \bar{\rho}_+ = \rho(z=+L/2) \\
        \bar{m}_- = m(z=-L/2) &\qquad& \bar{m}_+ = m(z=+L/2).
    \end{aligned}
    \label{eqn:slab_BC}
\end{equation}
As one sees, the confinement induces an interaction term $U_s$ composed of two contributions, one in $z=-L/2$ and one in $L/2$.
Minimizing the total energy with respect to the polarization field still leads to a fourth order differential equation :
\begin{equation}
    (1+K) m(z) - K_l m^{(2)}(z) + \beta m^{(4)}(z) = D_0.
\end{equation}
The solutions of this differential equation are written in the following form:
\begin{equation}\begin{split}
    m_1(z) = \frac{D_0}{1+K}
        &+ \bar{C}_1 \cos{(q_o z)} \cosh{(q_d z)} 
           + \bar{C}_2 \sin{(q_o z)} \cosh{(q_d z)} \\
        &+ \bar{C}_3 \cos{(q_o z)} \sinh{(q_d z)}
           + \bar{C}_4 \sin{(q_o z)} \sinh{(q_d z)}.
\end{split}
\label{eqn:slab_m}\end{equation}
Written in this manner, we underline the fact that the polarization $m(z)$ is odd (and the bound charge $\rho(z)=-\dot{m}(z)$ is even) for a vanishing $D_0$.\par
\begin{figure}[H]
    \centering 
    \includegraphics{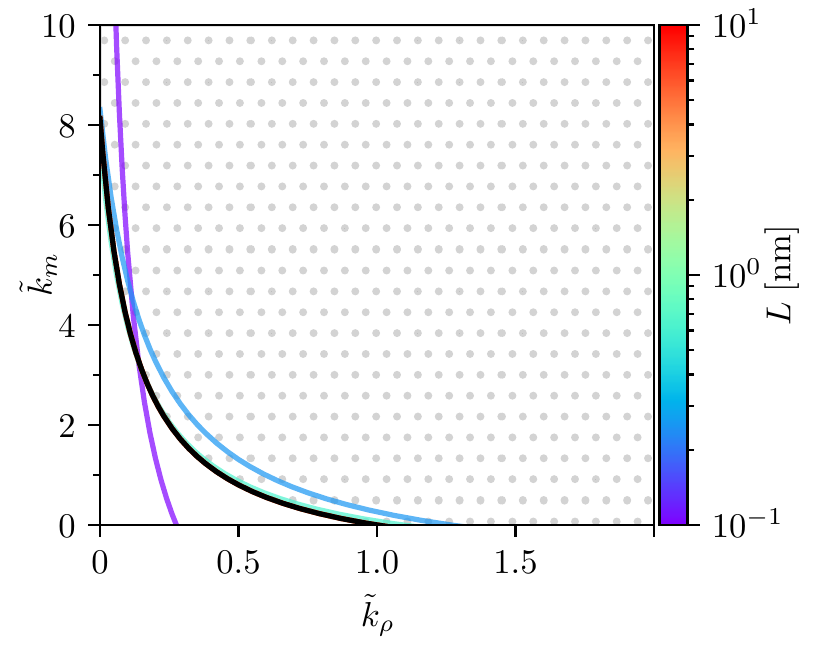}
    \caption{The curves are plotted for a null-Hessian matrix determinant of the functional energy and for $D_0=0$. The color of the curves is related to the distance $L$ between the two walls given on the colorbar.
    The black curve is drawn for a system with only one wall.}
    \label{SI-fig:slab_stability}
\end{figure}

Following the approach presented in the previous section, the solution of the differential equation (Eq.~\ref{eqn:slab_m}) is introduced into the functional energy $U$. 
One then gets an energy $U(\bar{C}_1, \bar{C}_2, \bar{C}_3, \bar{C}_4)$  which stationary point $(C_{1s}, C_{2s}, C_{3s}, C_{4s})$ can be determined by cancelling its gradient.
Results are given in the appendix.
Note that in the case where the external field $D_0=0$, the problem is much simpler.
Indeed, the system is then symmetrical according to $z$: $m(z) = -m(-z)$ which leads to having $\bar{C}_1=\bar{C}_4=0$.
The energy $U(\bar{C}_2, \bar{C}_3)$ is then a function of two variables which nature of the extremum can be assessed by studying the sign of the determinant of the associated Hessian matrix.
Figure~\ref{SI-fig:slab_stability} shows that as soon as the distance between the two walls is greater than $\sim \SI{1}{nm}$, one gets the same condition for the existence of a finite minimum as in the case where there is only one wall.
We do not consider cases where $L < \SI{1}{nm}$ as the continuous approximation of the confined water would then be unreasonable.
Thus the addition of the second wall has no significant effect on the nature of the extremum.
\par
\section{The details of molecular dynamics simulations}

We performed molecular dynamics simulations using GROMACS 2019.6 package.\cite{abraham2015}
Dispersion interactions are modelled using effective Lennard-Jones (LJ) potentials truncated at  $\SI{1.2}{nm}$ using a Verlet cutoff scheme. 
The truncated Lennard-Jones potential between two atoms separated by a distance $r_{ij}$ is
\begin{equation}
    V_{LJ}(\mathbf{r}_{ij}) = 
    \begin{cases}
        4\epsilon_{ij}\left(\left(\frac{\sigma_{ij}} {{r_{ij}}}\right)^{12} - \left(\frac{\sigma_{ij}}{{r_{ij}}}\right)^{6} \right) &\text{ for } r_{ij} \leq \SI{1.2}{nm} \\
        0 &\text{ for } r_{ij} > \SI{1.2}{nm}
    \end{cases}
    ,
\end{equation}
where $\epsilon_{ij}$ is the depth of the potential well (or dispersive energy) and $\sigma_{ij}$ is the distance at which the particle-particle potential energy is zero (or size of the particle).
Only atom pairs involving an oxygen atom interact through the Lennard-Jones potential. 
Associated parameters, $\epsilon_{\mathrm{X-O}}$ and  $\sigma_{\mathrm{X-O}}$, are given in the following Table~\ref{SI-table:LJ}.
\begin{table}[h]
    \centering
    \begin{tabular}{|l|l|l|l|c|}
    \hline
    X & Partial charge [e] & $\sigma_{\mathrm{X-O}}\;\mathrm{[nm]}$ & $\epsilon_{\mathrm{X-O}}\;\mathrm{[kJ/mol]}$ & Source \\  \hline
    H & 0.4238 & 0 & 0 & \\ \hline
    O & -0.8476 & 0.3166 & 0.650 & \\ \hline
    C & 0 & 0.319 & 0.392 & Werder et al.\cite{werder2003} \\  \hline
    B & 0.37 & 0.331 & 0.508 & Won and Aluru\cite{won2007} \\  \hline
    N & -0.37 & 0.326 & 0.628& Won and Aluru\cite{won2007} \\  \hline
    \end{tabular}
    \caption{Partial charges and Lennard-Jones parameters for the interaction between the atoms of the wall X and the oxygen of the water molecules. }
    \label{SI-table:LJ}
\end{table}\par

The Coulomb force is treated using a real-space cutoff at $\SI{1.2}{nm}$ and particle mesh Ewald summation (pseudo-2D particle mesh Ewald summation for slab geometry\cite{yeh1999}). 
We use the three-site model SPC/E for water molecules.
Simulations performed in the NVT ensemble are thermalized at $\SI{300}{K}$ with the v-rescale thermostat \cite{bussi2007} (relaxation time $\tau=\SI{0.5}{ps}$). 
Furthermore, all MD simulations have performed with a time step equal to $\SI{2}{fs}$.\par

\subsection{Periodic water box}

\begin{figure}[H]
    \centering 
    \includegraphics{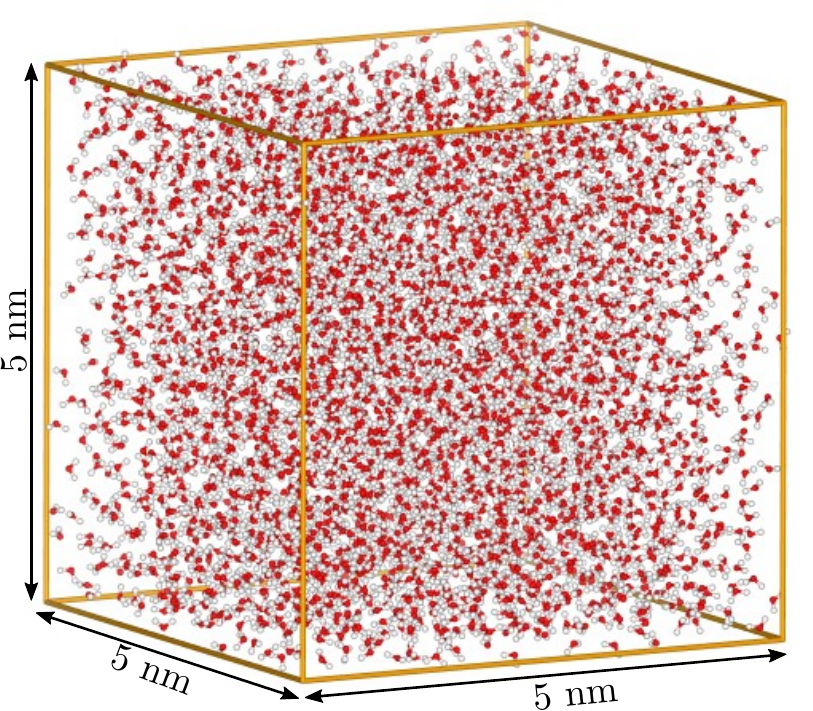}
    \caption{The simulated system after the thermalization in NVT ensemble: a $\SI{5}{nm}$ side cube, periodic along the 3 directions of space and filled with 4055 SPC/E water molecules. }
    \label{SI-fig:system_waterbox}
\end{figure}\par

The simulated system is a $5\times5\times\SI{5}{nm^3}$ cube containing 4055 SPC/E water molecules and with periodicity following the 3 directions of space. 
A first simulation carried out in an NVT canonical ensemble and of short duration allows to ensure the thermalization of the system.
Then we let the system adjust the volume of the box at constant pressure equal to $\SI{1}{bar}$ with a Berendsen barostat \cite{berendsen1984} ($\tau=\SI{2.0}{ps}$) in order to reach a density close to bulk water ($\sim \SI{998}{kg/m^3}$) thanks to a second simulation performed in a NPT ensemble with a long range dispersion corrections for energy and pressure.
At the end of this relaxation, the side of the box is slightly different and equal to $\SI{4.963}{nm}$.
Finally, we carry out long-term simulations ($\SI{2}{ns}$) in a canonical ensemble from which we compute H-H, O-H and O-O structure factors and deduce the susceptibility $\hat{\chi}_\parallel(q)$ with the classical approximation for the fluctuation dissipation theorem (see black curves on Figure~\ref{SI-fig:chi_q_fitting}).\cite{bopp1996}\par
Figure~\ref{SI-fig:chi_q_fitting} shows the two strategies to fit the dielectric susceptibility of bulk water with a 4 order model:
\begin{equation}
    \hat{\chi}_\parallel(q)= \frac{1}{1 + K +K_l q^2 + \beta q^4}.
\end{equation}
The first one aims at fitting the maximum of the susceptibility while the second one focuses on fitting the width of the susceptibility normalized to 1 (the aspect ratio) bearing in mind that in both cases $\hat{\chi}_\parallel(q=0)$ is set by the value of the relative dielectric permittivity of the water SPC/E $\epsilon_r = (1-\hat{\chi}_\parallel(q=0))^{-1} \sim 71$.
\begin{figure}[H]
    \centering 
    \includegraphics{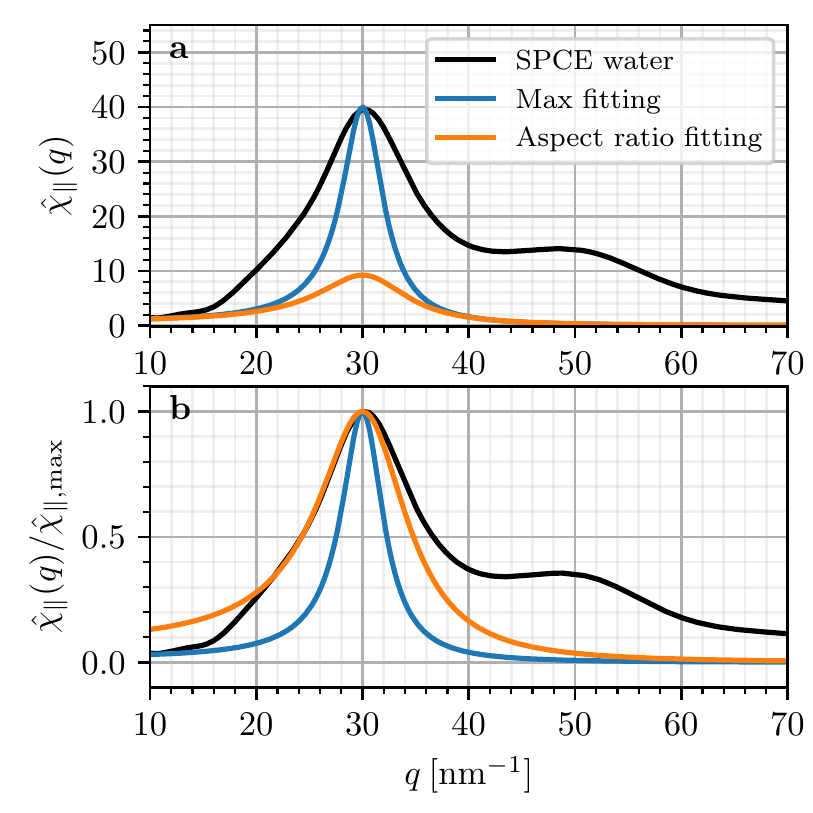}
    \caption{Susceptibility $\chi(q)$ (\textbf{a}) and normalized susceptibility $\chi(q)/\chi_{max}$ (\textbf{b}) computed from pure SPC/E water molecular dynamic simulation (black curve). The blue curve arises from the fit of the maximum of the first component of the susceptibility while the orange curve shows the result of the fit of the aspect ratio. }
    \label{SI-fig:chi_q_fitting}
\end{figure}\par

\subsection{Water confined in slab geometry}
\begin{figure}[H]
    \centering 
    \includegraphics{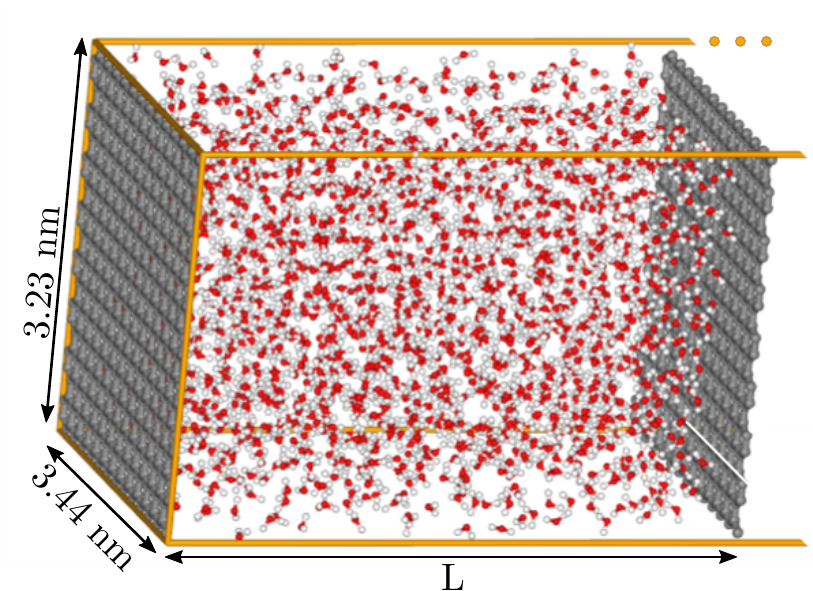}
    \caption{The simulated system after the thermalization in NVT ensemble: SPC/E water molecules confined between two graphene sheets.}
    \label{SI-fig:system_slab}
\end{figure}\par

We performed molecular dynamics simulations of SPC/E water confined in a slab geometry.
The walls, perpendicular to the $z$ direction, are made up of atoms which positions are fixed and arranged in a hexagonal lattice.
$L$ is the distance between the two walls.
The simulation box is extended in the z-direction until it reaches a length equal to $3 L$.
Thus, even if we considered a periodic system in the 3 directions of space, the periodized slab system is separated by a $2L$-thick void layer along the z axis.\par

In our study we considered two types of hydrophobic walls: graphene and a hBN wall.
The dispersion interactions between the wall and the water molecules are modelled using the truncated Lennard-Jones potential between the atoms of the wall and the oxygen atoms of the water molecules. 
We recall that the parameters of this potential are given in the Table~\ref{SI-table:LJ}.
The walls are electrically neutral overall but, in the case of hBN, the atoms carry partial charges leading to an electrostatic interaction with the water molecules.\par

As for the previous system, we conducted a first simulation in a NVT canonical ensemble and of short duration that allows to ensure the thermalization of the system.
Then, we check that the number N of water molecules introduced between the two walls, separated by a distance $L$, leads to a density close to bulk water ($\sim \SI{1000}{kg/m^3}$) with a second simulation performed in a NPT ensemble with a semi-isotropic Parrinello-Rahman barostat \cite{bussi2009} ($\tau=\SI{2.0}{ps}$).
Finally, we carry out several long-term simulations ($\SI{60}{ns}$) in a canonical ensemble with homogeneous external electric fields aligned along the z direction.
This procedure was performed for (i) graphene/graphene slab and (ii) hBN/graphene slab (see Figure~\ref{SI-fig:simulations_5nm}) 
In both case, the two walls are separated by $\SI{5}{nm}$ and an electric field intensity ranging from 0 to $\SI{2}{V/nm}$.
Besides, we have computed simulations with two separated graphene walls with a length L ranging from 1 to $\SI{10}{nm}$ with $E_{ext}=0$ and $\SI{0.5}{V/nm}$.\par

From the molecular trajectories, we compute the charge density $\rho(z)$ as a histogram of the point charge distribution with a step of $\SI{0.01}{nm}$ along $z$.
Then, the polarization field $m(z)$ calculate by performing a numerical derivative of the charge density, $m(z) = - d \rho(z)/dz$.
Finally, the susceptibility is computed from the polarization field of two simulations in which the amplitude of the external field is different by a value $\delta D_0=\SI{0.5}{V/nm}$ :
\begin{equation}
    \chi(z, D_0) = \frac{m(z, D_0 + \delta D_0) - m(z, D_0)}{\delta D_0}
\end{equation}\par

\subsubsection{Validation  of the fitting strategy of the bulk susceptibility $\hat{\chi}_\parallel(q)$} 
\begin{figure}[H]
    \centering 
    \includegraphics{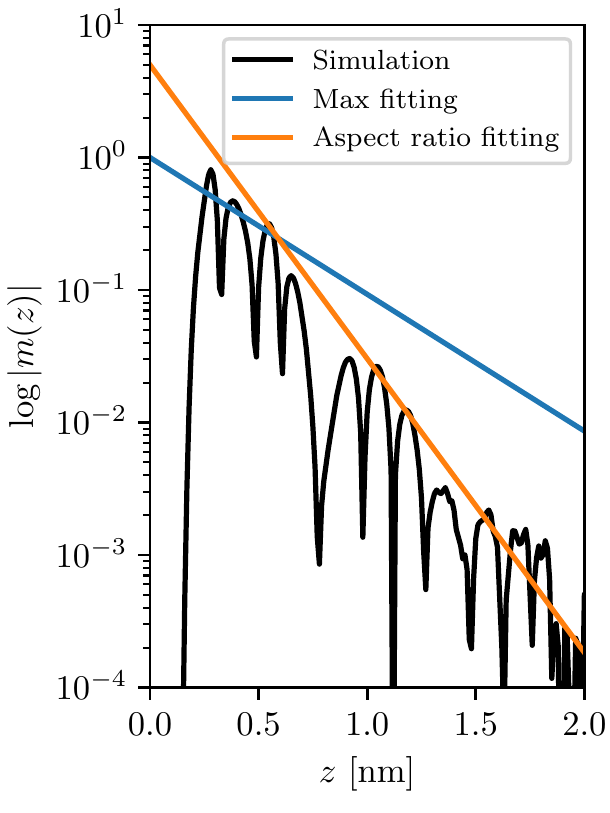}
    \caption{The black curve is the polarisation field computed from MD simulation of water in the vicinity of graphene layer at vanishing $D_0$. 
    Color curves have the equation $m(z) = A e^{-q_d z}$ where $q_d$ is computed from the fitting of the bulk susceptibility $\hat{\chi}(q)$.
    The blue one arises from the fit of the maximum of the first component of the susceptibility while the orange curve shows the result of the fit of the aspect ratio.
    The parameter $A$ is chosen to match the back curve.}
    \label{SI-fig:chi_q_fitting}
\end{figure} \par

\subsubsection{Electrostatic field for large slab geometry}
\begin{figure}[H]
    \centering 
    \includegraphics{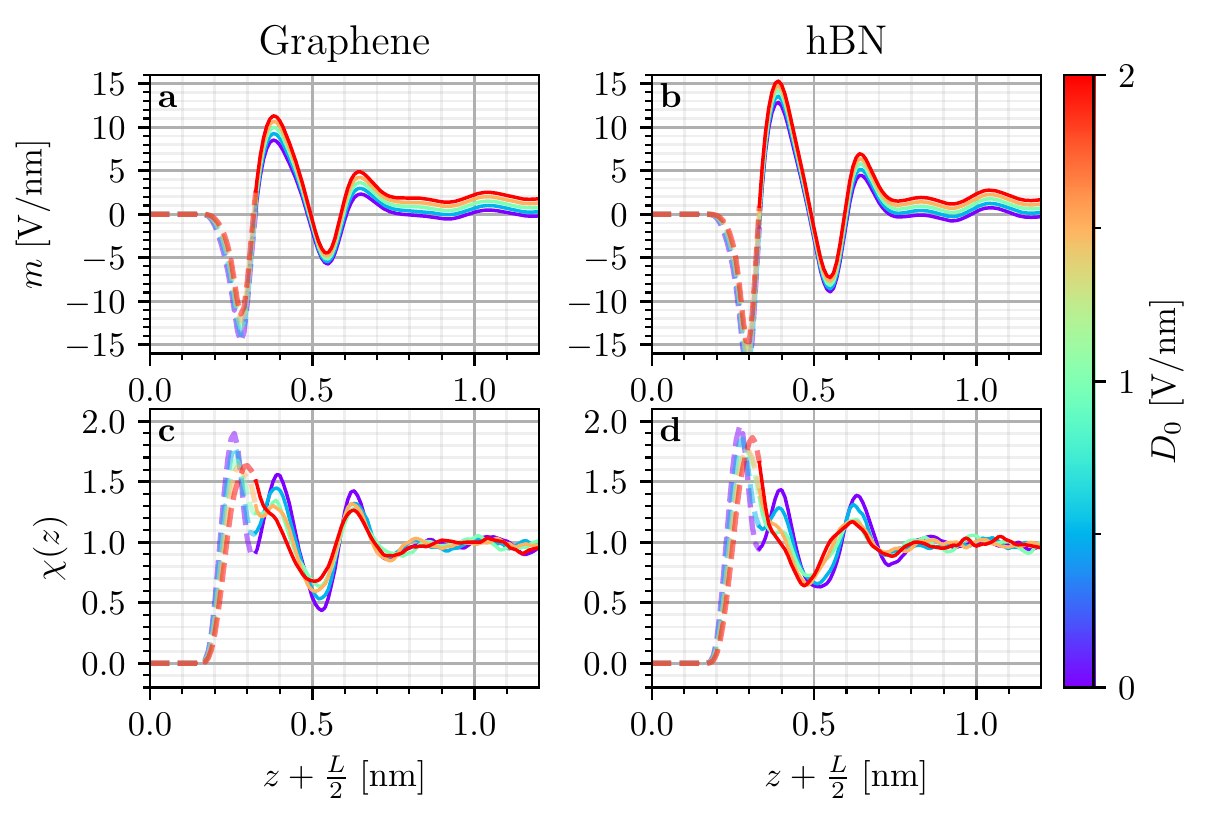}
    \caption{Polarization $m(z)$ (top panel) and susceptibility $\chi(z)$ (bottom panel) from molecular dynamics simulations of water near a graphene (left panel) or hBN layer (right panel) for different external field strengths (see color code).
    The part of the curves for $z<z_0$ are shown as dotted lines.}
    \label{SI-fig:simulations_5nm}
\end{figure}

The simulations for which the results are shown in figures~\ref{SI-fig:simulations_5nm} were performed with a $\SI{5}{nm}$ distance between the two graphene walls.
This value is much larger than the characteristic decay length $\lambda_d$.
Therefore, the water structuration in the vicinity of one wall is not affected by the presence of the second wall.
We then choose to display the results for the left wall only.\par
Figure~\ref{SI-fig:simulations_5nm} shows the polarization and susceptibility in such a system for increasing values of the external field $D_0$ up to $\SI{2}{V/nm}$. 
Considering the variations in amplitude of the susceptibility for interfacial water, one can estimate that the regime is linear up to $D_0 \sim \SI{2}{V/nm}$.
This value is however lower than the one reported in previous work.\cite{bonthuis2012} \HB{Cette valeur a aussi été estimée dans les Si de ... ???}

 \par

\subsubsection{Smearing the surface}
\begin{figure}[H]
    \centering
    \includegraphics{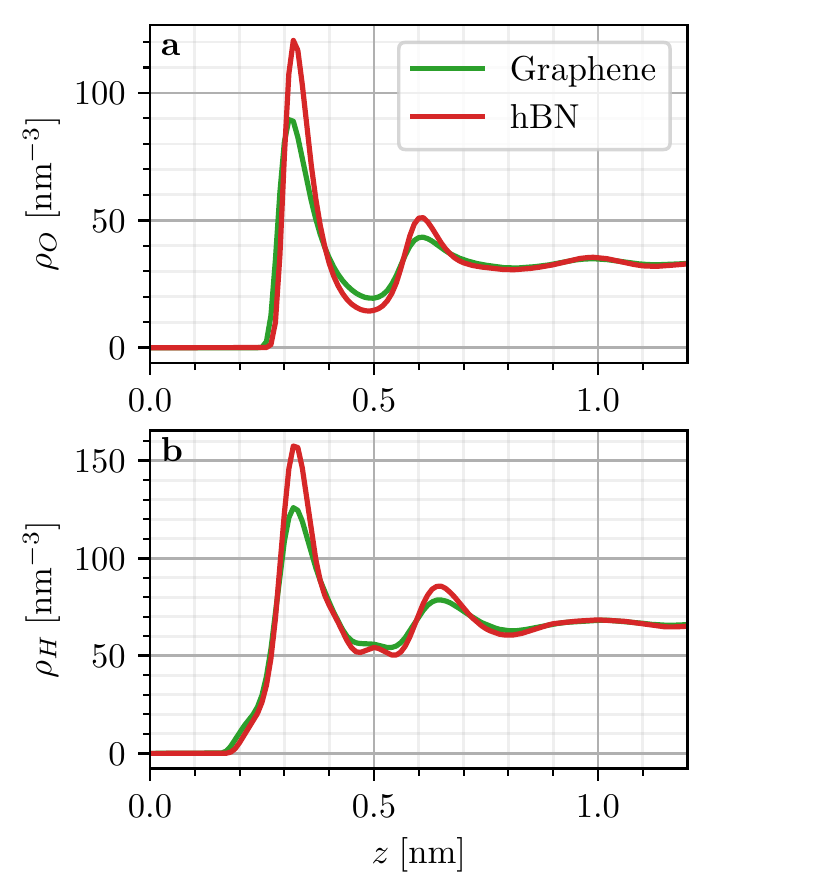}
    \caption{Oxygen $\rho_O(z)$ (top panel) and Hydrogen $\rho_H(z)$ density  near a graphene or a hBN layer.}
    \label{SI-fig:OHdensity}
\end{figure}

\begin{figure}[H]
    \centering
    \includegraphics{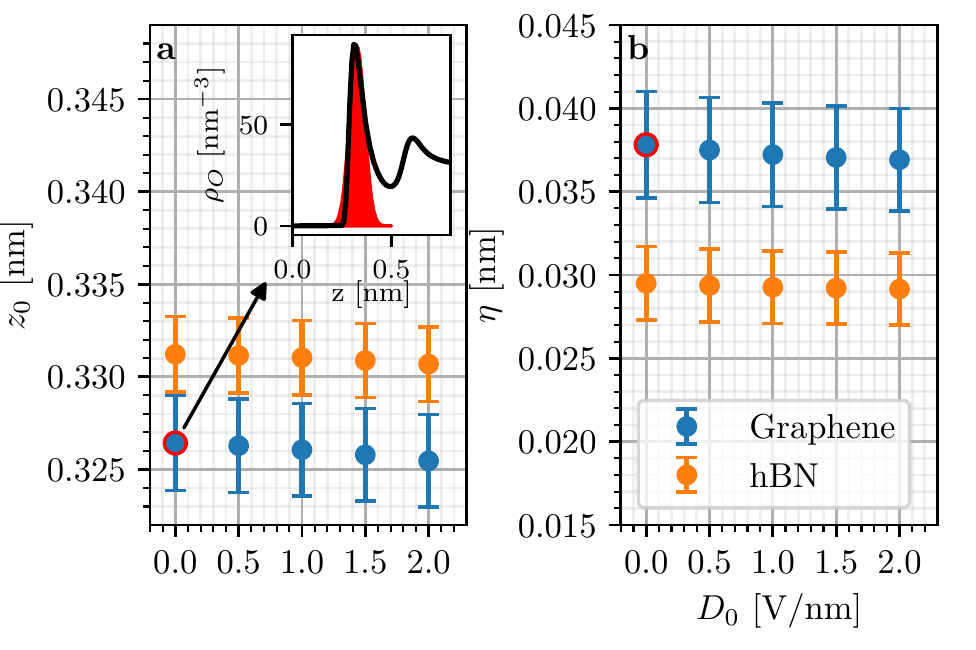}
    \caption{Smearing parameters $z_0$ (a) and the associated standard deviation $\eta$ (b) as a function of the external field and for different types of walls: graphene (blue), hBN (orange) and hBN without partial charges (green). 
    The inset illustrates the Gaussian fit of the first oxygen layer from which the parameters $z_0$ and $\eta$ are derived.}
    \label{SI-fig:smearing}
\end{figure}

The Figure~\ref{SI-fig:OHdensity} represents the density of Oxygen $\rho_O$ and Hydrogen $\rho_H$  in the vicinity of graphene (green curve) and hBN (red curve) layers.
The surface induces a layering of the fluid that is more pronounced for hBN as for graphene which is indicated by a higher and narrower first hydration peak for this surface.
In order to take into account that the numerical interface is not 2D but is associated with a small width, we convolute the theoretical expressions of polarization and susceptibility with a Gaussian
\begin{equation}
    G(z-z_0) = \frac{e^{-z^2/2\eta^2}}{\eta\sqrt{2\pi}}.
\end{equation}
The position $z_0$ and the width $\eta$ of the Gaussian are determined for each surface by fitting the first peak of the Oxygen density with $G(z-z_0)$.
The inset of Fig~\ref{SI-fig:smearing} shows the Gaussian G (full red zone) and the simulated density for oxygen in the vicinity of graphene. 
The parameters $\eta$ and $z_0$ do not vary with the application of an excitation field $D_0$ as shown in Fig~\ref{SI-fig:smearing}.
\par

\par

\subsubsection{Molecular dynamics simulations as function of the slab width $L$}
\begin{figure}[H]
    \centering
    \includegraphics{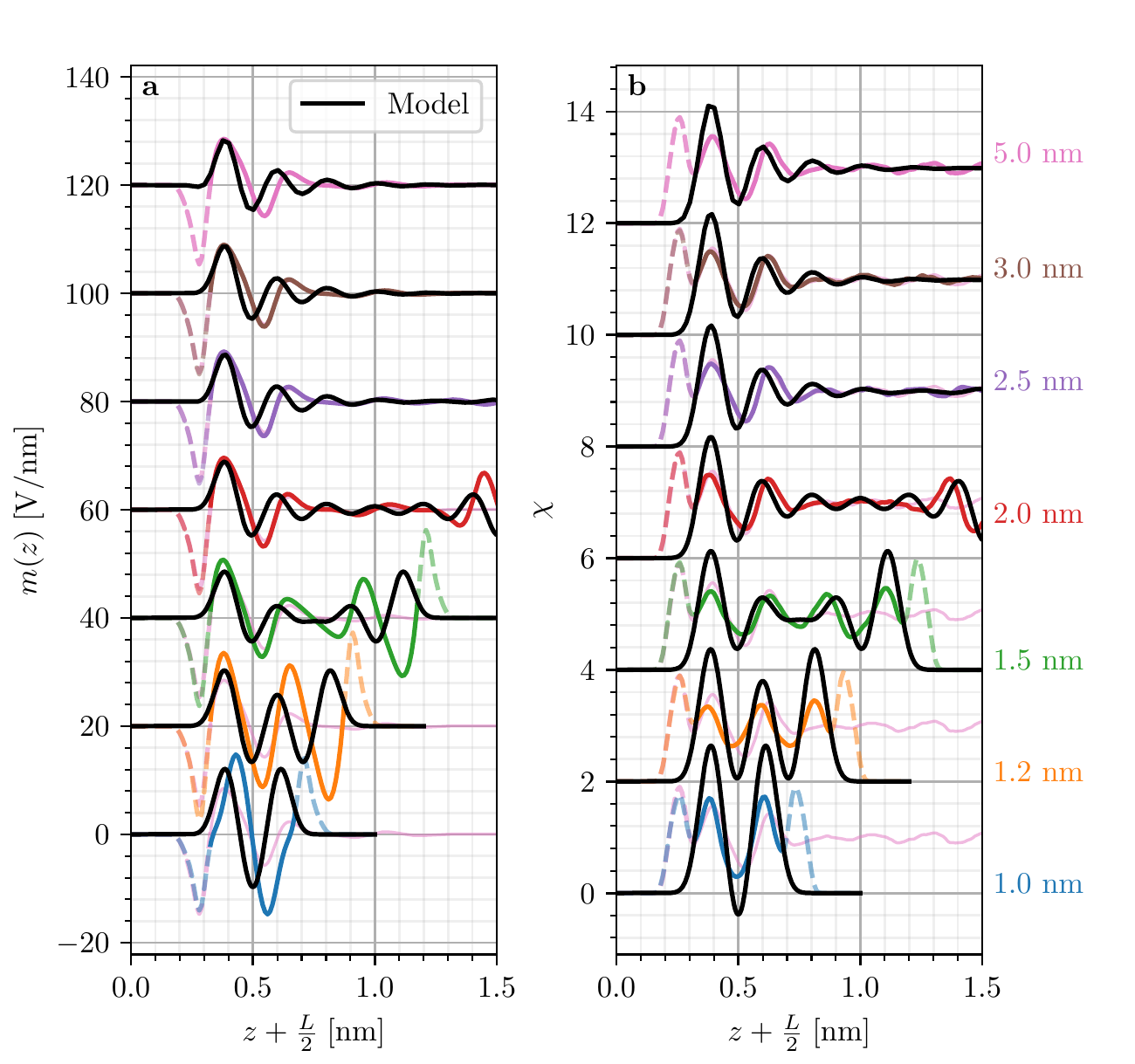}
    \caption{Polarization $m(z)$ (a) and susceptibility $\chi(z)$ (b) from molecular dynamics simulations of water in a slab geometry with graphene on both sides of the system and without external field.
    The width $L$ of the geometry is given on the right side of the figure.
    The colored curves, coming from MD simulations, are shown as dashed line for $z < -L/2 + z_0$ and $z > L/2 - z_0$.
    Curves have been shifted for sake of clarity. }
    \label{SI-fig:sim_L}
\end{figure}

The theoretical model that we have developed can be used to determine the profile of the polarization and the susceptibility of a nanometric water layer of varying thickness between graphene surfaces. The Figure~\ref{SI-fig:sim_L} shows $m(z)$ (a) and $\chi(z)$ for water thickness going from $\SI{5}{nm}$ to $\SI{1}{nm}$. 
The theoretical model allows to retrieve the organization in successive layers of over ($\chi(z) \gg \chi_b$) and underresponding ($\chi(z) \ll \chi_b$) region in interfacial water. \par

\par

\subsubsection{Physical origin of a non-zero $\tilde{k}_\rho$}
In this section, we probe the influence of partial charge $q$ and dispersion energy $\epsilon$ of the surface's atoms on the dielectric properties of interfacial water.
We consider artificial surfaces for which one of these parameters has been modified with respect to graphene or hBN.

\begin{figure}[H]
    \centering
    \includegraphics{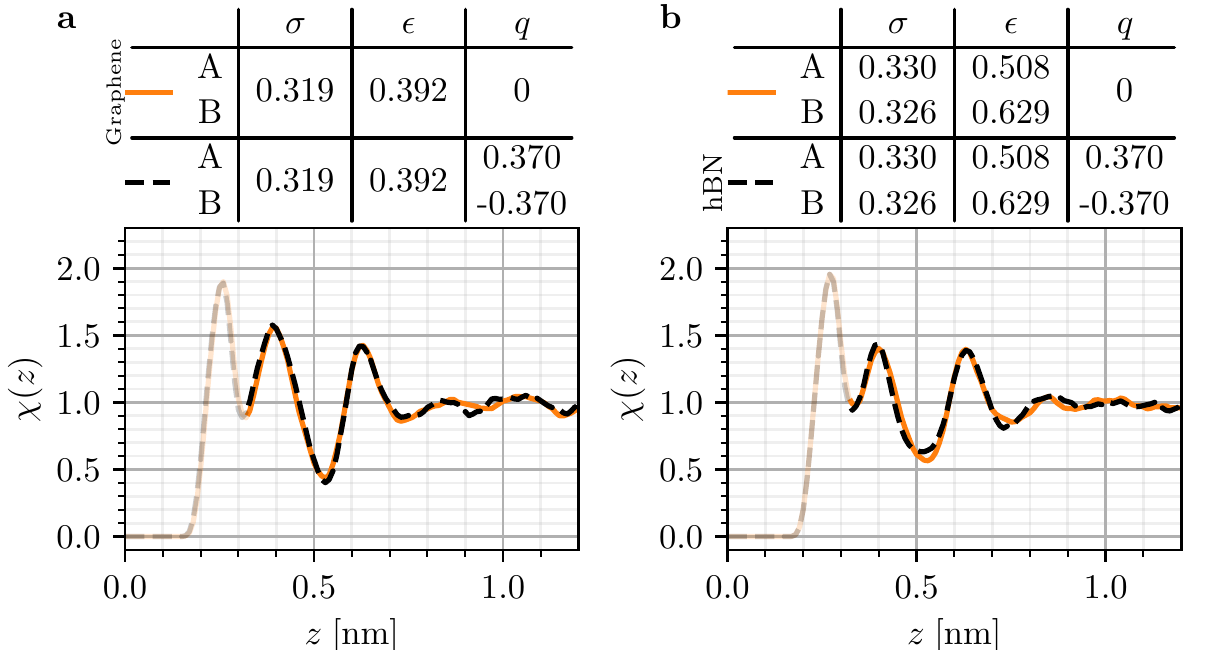}
    \caption{Dielectric susceptibility of water in the vicinity of graphene (a) or hBN (b) layer without and with partial charges $q$.
    $\sigma$ is given in nm, $\epsilon$ in $\mathrm{kJ/mol}$ and the partial charge $q$ in eV.
    The curves are made transparent for $z<\SI{0.33}{nm}$.
    }
    \label{SI-fig:nonzerokm_charge}
\end{figure}

\paragraph{Impact of partial charges}
First, we study the influence of partial charge of surface's atoms by considering a 'charged' graphene as well as 'non-charged' hBN.
The parameters of these two surfaces are given in tables~\ref{SI-fig:nonzerokm_charge}a and \ref{SI-fig:nonzerokm_charge}b.
From those custom MD simulation, we show that the dielectric susceptibility of interfacial water is the same for 'charged' graphene and graphene. 
It is also the same for 'non-charged' hBN and hBN.
We conclude that it is not the partial charge of the surface's atoms that trigger the differences of the dielectric susceptibility of graphene and hBN interfacial water. 
\par

\begin{figure}[H]
    \centering
    \includegraphics{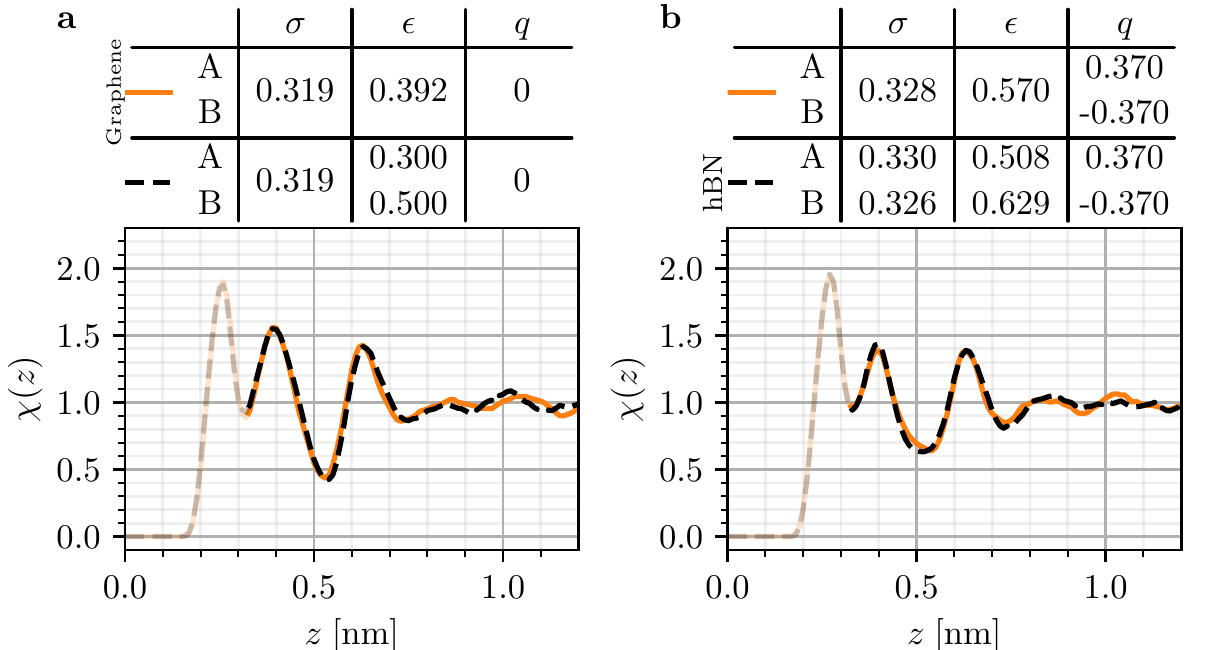}
    \caption{Focus on the impact of a bi-atomic wall on the dielectric susceptibility of water.
    The orange curves are derived from a mono-atomic wall while the black dashed curves result from a bi-atomic wall. 
    $\sigma$ is given in nm, $\epsilon$ in $\mathrm{kJ/mol}$ and the partial charge $q$ in eV.
    The curves are made transparent for $z<\SI{0.33}{nm}$.}
    \label{SI-fig:nonzerokm_heterogeneity}
\end{figure}

\paragraph{Impact of heterogeneity}
Secondly, we evaluate the impact of the bi-atomic nature of hBN on the susceptibility. 
To do so, we consider a 'bi-atomic' like graphene surface associated with LJ parameters given in Table~\ref{SI-fig:nonzerokm_heterogeneity}a.
The two artificial atoms have the same size $\sigma$ as carbon's in graphene and do not carry partial charges.
However, they are associated with two different $\epsilon$ which average is equal to the carbon's dispersion energy in graphene.
In contrast, we simulate a 'mono-atomic' like hBN surface (see Table~\ref{SI-fig:nonzerokm_heterogeneity}b).
This surface is composed of two atoms carrying opposite partial charges equal to $\pm \SI{0.37}{e}$.
These two atoms are associated with the same LJ parameter set. 
The size and the dispersion energy are equal to the average of those for Boron and Nitride.
From those custom MD simulations, we show that the dielectric susceptibility of interfacial water is the same for 'bi-atomic' like graphene and graphene as well as 'mono-atomic' like hBN and hBN.
This aspect has no influence on the dielectric susceptibility of interfacial water. 
\par

\begin{figure}[H]
    \centering
    \includegraphics{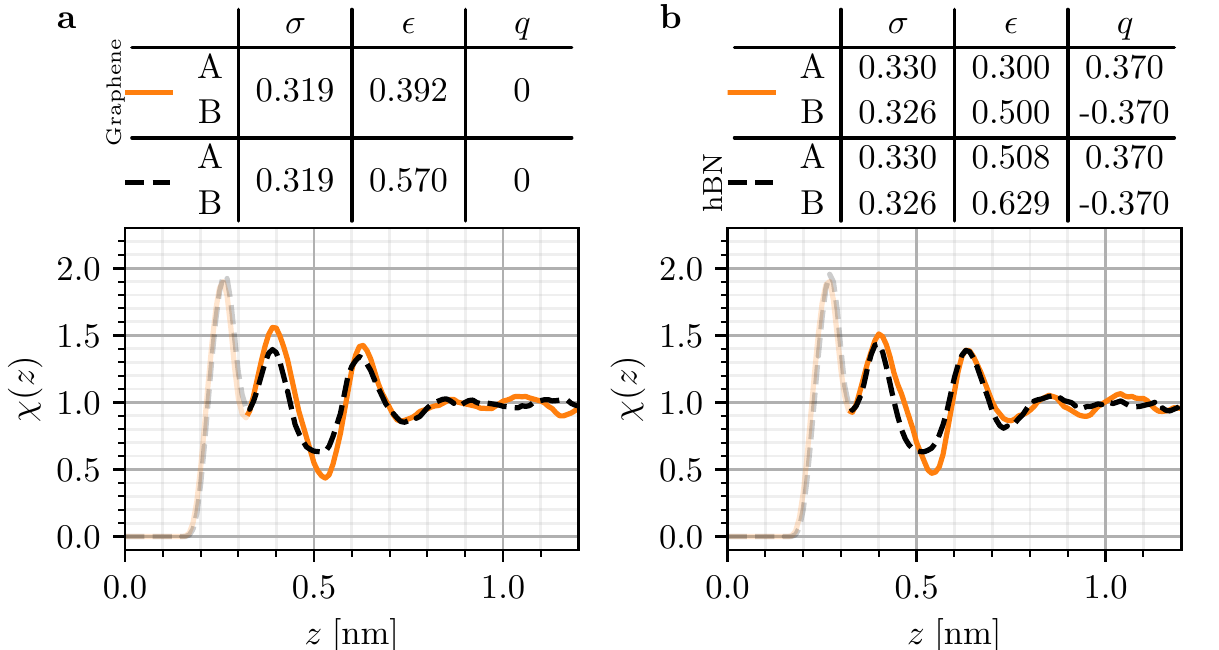}
    \caption{Dielectric susceptibility of water in the vicinity of a surface when the Lennard-Jones interaction between the water and this surface is deeper (from orange curves to black dashed curves).
    $\sigma$ is given in nm, $\epsilon$ in $\mathrm{kJ/mol}$ and the partial charge $q$ in eV.
    The curves are made transparent for $z<\SI{0.33}{nm}$.}
    \label{SI-fig:nonzerokm_LJdepth}
\end{figure}

\paragraph{Average dispersion energy influence}
Finally, we increase the dispersion energy of carbon atoms in the graphene surface (see Table~\ref{SI-fig:nonzerokm_LJdepth}a).
In contrast, we decrease the average dispersion energy for hBN by keeping the ratio between the two atoms (see Table~\ref{SI-fig:nonzerokm_LJdepth}b).
We can see that the greater the average dispersion energy, the smaller the susceptibility amplitude.

\par

The susceptibility of interfacial water for graphene surface shows a higher amplitude than the one for hBN surface. 
Thanks to this investigation, we conclude that this difference is triggered by the average dispersion energy which is smaller for graphene.\par

\appendix
\section{Appendix : Stationary point in the case of a slab geometry}
\begin{align}
    C_{1s} &= \left(m_0-\frac{D_0}{1+K}\right) \frac{a_1}{b+c}\\
    C_{2s} &= \rho _0 \frac{a_2}{b-c}\\
    C_{3s} &= \rho _0 \frac{a_3}{b-c}\\
    C_{4s} &= \left(m_0-\frac{D_0}{1+K}\right) \frac{a_4}{b+c}
\end{align}
with
\begin{align}
    a_1 &= 
    \begin{aligned}[t]
        2 k_m  \epsilon _0 \left(q_d^2+q_o^2\right) 
        &\left(\sinh \left(\frac{L q_d}{2}\right) \left(\epsilon _0 k_{\rho } q_o \left(q_d^2+q_o^2\right){}^2 \cos \left(\frac{L q_o}{2}\right)+(K+1) \left(q_d-q_o\right) \left(q_d+q_o\right) \sin \left(\frac{L q_o}{2}\right)\right) \right.\\
        & + \left.  \cosh \left(\frac{L q_d}{2}\right) \left(\epsilon _0 k_{\rho } q_d \left(q_d^2+q_o^2\right){}^2 \sin \left(\frac{L q_o}{2}\right)+2 (K+1) q_o q_d \cos \left(\frac{L q_o}{2}\right)\right)\right)
    \end{aligned}\\
    a_2 &=
    \begin{aligned}[t]
        2 \epsilon _0 k_{\rho } \left(q_d^2+q_o^2\right)^2
        & \left(\sinh \left(\frac{L q_d}{2}\right) \left(\epsilon _0 k_m \left(q_d^2+q_o^2\right) \cos \left(\frac{L q_o}{2}\right) + (K+1) q_o \sin \left(\frac{L q_o}{2}\right)\right)\right. \\
        & + \left. (K+1) q_d \cosh \left(\frac{L q_d}{2}\right) \cos \left(\frac{L q_o}{2}\right)\right)
    \end{aligned}\\
    a_3 &=
    \begin{aligned}[t]
        -2 \epsilon_0 k_{\rho } \left(q_d^2+q_o^2\right)^2
        &\left(\cosh \left(\frac{L q_d}{2}\right) \left(\epsilon _0 k_m \left(q_d^2+q_o^2\right) \sin \left(\frac{L q_o}{2}\right)-(K+1) q_o \cos \left(\frac{L q_o}{2}\right)\right)\right. \\
        & + \left. (K+1) q_d \sinh \left(\frac{L q_d}{2}\right) \sin \left(\frac{L q_o}{2}\right)\right)
    \end{aligned}\\
    a_4 &= 
    \begin{aligned}[t]
        - 2 k_m \epsilon _0 \left(q_d^2+q_o^2\right) 
        &\left(\sinh \left(\frac{L q_d}{2}\right) 
        \left(
            \epsilon _0 k_{\rho } q_d \left(q_d^2+q_o^2\right)^2 \cos \left(\frac{L q_o}{2}\right)
            - 2 (K+1) q_o q_d \sin \left(\frac{L q_o}{2}\right)
        \right)\right.\\
        &+ \left. \cosh \left(\frac{L q_d}{2}\right)
        \left(
            -\epsilon _0 k_{\rho } q_o \left(q_d^2+q_o^2\right)^2 \sin \left(\frac{L q_o}{2}\right)
            + (K+1) \left(q_d-q_o\right) \left(q_d+q_o\right) \cos \left(\frac{L q_o}{2}\right)
        \right)\right)
    \end{aligned}
\end{align}
and
\begin{align}
    b&=
    \begin{aligned}[t]
        -&2 (K+1) \epsilon _0 q_d q_o \left(q_d^2+q_o^2\right) \cos \left(L q_o\right) \left(k_{\rho } \left(q_d^2+q_o^2\right)-k_m\right)\\
        +& q_d \sin \left(L q_o\right) \left(\epsilon _0^2 k_m k_{\rho } \left(q_d^2+q_o^2\right){}^3-(K+1)^2 \left(q_d^2-3 q_o^2\right)\right)
    \end{aligned}
    \\
    c&=
    \begin{aligned}[t]
        &2 (K+1) \epsilon _0 q_d q_o \left(q_d^2+q_o^2\right) \cosh \left(L q_d\right) \left(k_{\rho } \left(q_d^2+q_o^2\right)+k_m\right)\\
        & q_o \sinh \left(L q_d\right) \left(\epsilon _0^2 k_m k_{\rho } \left(q_d^2+q_o^2\right){}^3+(K+1)^2 \left(3 q_d^2-q_o^2\right)\right)
    \end{aligned}
\end{align}

\medskip
\bibliographystyle{unsrt}
\bibliography{references}